\documentclass[a4paper, 12pt]{article}
\usepackage{epsfig}
\usepackage{amssymb}
\usepackage{amsthm}
\usepackage[cmex10]{amsmath}
\usepackage{amsmath}
\usepackage{amsfonts}
\usepackage{amssymb}
\usepackage{bbm}
\usepackage{multicol}
\usepackage{graphicx}
\usepackage[left=1.5cm,right=2.5cm]{geometry}
\usepackage{enumitem}
\usepackage{natbib}
\usepackage[colorlinks=true,linkcolor=blue, citecolor=blue, urlcolor=blue]{hyperref}
\usepackage{epsfig}
\usepackage[dvipsnames]{xcolor}
\usepackage{epstopdf}
\usepackage{caption}
\usepackage[textfont=footnotesize,justification=
 centering]{subcaption}
\usepackage{multirow}
\usepackage{arydshln}
\usepackage{fancyhdr}
\usepackage{appendix}
\usepackage[hang,flushmargin]{footmisc}
\usepackage{titling}
\usepackage{soul}
\usepackage{framed}
\usepackage{color}

\newtheorem*{theorem-non}{Theorem}
\newtheorem{theorem}{Theorem}
\newtheorem{definition}[theorem]{Definition}
\newtheorem{proposition}[theorem]{Proposition}

\newtheorem{corollary}[theorem]{Corollary}

\newtheorem{example}[theorem]{Example}

\colorlet{shadecolor}{yellow}
\input{tcilatex}
\begin{document}

\title{Forward Rank-Dependent Performance Criteria: Time-Consistent
Investment Under Probability Distortion}
\author{Xue Dong He\thanks{%
Department of Systems Engineering and Engineering Management, The Chinese
University of Hong Kong. Email: xdhe@se.cuhk.edu.hk} \and Moris S. Strub%
\thanks{%
Department of Systems Engineering and Engineering Management, The Chinese
University of Hong Kong. Email: msstrub@se.cuhk.edu.hk} \and Thaleia
Zariphopoulou\thanks{%
Departments of Mathematics and IROM, The University of Texas at Austin and
the Oxford-Man Institute, University of Oxford. Email:
zariphop@math.utexas.edu}}
\maketitle

\begin{abstract}
We introduce the concept of forward rank-dependent performance processes, extending the original notion to forward criteria that incorporate probability distortions. A fundamental challenge is how to reconcile the time-consistent nature of forward performance criteria with the time-inconsistency stemming from probability distortions.  For this, we first propose two distinct definitions, one based on the preservation of performance value and the other on the time-consistency of policies and, in turn, establish their equivalence. We then fully characterize the viable class of probability distortion processes, providing a “bifurcation”-type result. Specifically, it is either the case that the probability distortions are degenerate in the sense that the investor would never invest in the risky assets, or  the marginal probability distortion equals to a normalized power of the quantile function of the pricing kernel. We also characterize the optimal wealth process, whose structure motivates the introduction of a new, “distorted” measure and a related market. We then build a striking correspondence between the forward rank-dependent criteria in the original market and forward criteria without probability distortions in the auxiliary market. This connection also provides a direct construction method for forward rank-dependent criteria. A byproduct of our work are some new results on the so-called dynamic utilities and on time-inconsistent problems in the classical (backward) setting.  
\end{abstract}

{\textbf{\newline
Keywords:} forward criteria, rank-dependent utility, probability distortion,
time-consistency, portfolio selection, inverse problems}

\newpage

\section{Introduction}

\label{sec:Introduction} In the classical expected utility framework, there are three fundamental modeling ingredients, namely, the model, the trading horizon and the risk preferences, and all are chosen at initial time.
Furthermore, both the horizon and the risk preferences are set exogenously
to the market. In most cases, the Dynamic Programming Principle (DPP) holds
and provides a backward construction of the solution. This yields
time-consistency of the optimal policies and an intuitively pleasing
interpretation of the value function as the intermediate indirect utility. There
are, however, several limitations with this setting.

It is rarely the case that the model is fully known at initial time. Model
mis-specification and model decay occur frequently, especially as the
investment time increases. Even if a family of models is assumed, instead of
a single model, and robust control criteria are used, still the initial
choice of this family of models could turn out to be inaccurate quite fast.
This is also the case when filtering is incorporated, as it is based on the
dynamics of the observation process which, however, can be wrongly
pre-chosen. In addition, the trading horizon is almost never fixed, not even
fully known at the beginning of an investment period. It may change,
depending on upcoming (even unforeseen) opportunities and/or changes of risk preferences. Finally, it might be difficult to justify that one knows his utility far ahead in the future. It is more natural to know how one feels towards uncertainty for the immediate future, rather than for instances in the distant one (see, for example, the old note of Fischer Black, \cite{Black68}).

Some of these limitations have been successfully addressed. For example,
dynamic model correction is central in adaptive control where the model is
revised as soon as new incoming information arrives and, in turn,
optimization starts anew for the remaining of the horizon. Flexibility with
regards to horizons has been incorporated by allowing for rolling horizons
from one (pre-specified horizon end to the next). Risk preferences have been
also considered in more complex settings like recursive utilities, which are
modeled through a ``utility generator" that dictates a more sophisticated
backward evolution structure.

Nevertheless, several questions related to genuinely dynamic revision of
preferences and of the model, time-consistency across interlinked investment
periods as well as under model revisions, endogenous versus exogenous
specification of modeling ingredients, and others remain open. A
complementary approach that seems to accommodate some of the above
shortcomings is based on the so-called forward performance criteria. These
criteria are progressively measurable processes that, compiled with the
state processes along admissible controls, remain super-martingales and
become martingales at candidate optimal policies. In essence, forward
criteria are created by imposing the DPP forward, and not backwards, in
time. As a result, they adapt to the changing market conditions, do not rely
on an a priori specification of the full model, and accommodate dynamically
changing horizons. They produce endogenously a family of risk preferences
that follow the market in ``real-time" and, by construction, preserve
time-consistency across all times.

Forward criteria were introduced by \cite%
{MusielaZariphopoulou06,MusielaZariphopoulou08,
MusielaZariphopoulou09,MusielaZariphopoulou10a,
MusielaZariphopoulou10b,MusielaZariphopoulou11} and, further studied, among
others, in \cite{Zitkovic09}, \cite{ZariphopoulouZitkovic10}, \cite{ElKarouiMohamed13}, \cite{BernardKwak16}, \cite{ShkolnikovSircarZariphopoulou16},\linebreak \cite{ChoulliMa17}, \cite{ LiangZariphopoulou17} and \cite{ChongHuLiangZariphopoulou18}. More recently, they have also been considered in discrete-time by \cite%
{AngoshtariZariphopoulouZhouXX} and\linebreak \cite{StrubZhouXX}, applied to problems
arising in insurance by \cite{ChongXX}, extended to settings with model
ambiguity in \cite{KaellbladOblojZariphopoulou18} and \cite{ChongLiangXX},
and to optimal contract theory \cite{NadtochiyZariphopoulouXX}.

The associated optimization problems are ill-posed, as one specifies the
initial condition and solves the problem forward in time. In Ito markets, a
stochastic PDE for the forward criterion was derived in \cite{MusielaZariphopoulou10b}, while in \cite{LiangZariphopoulou17} a connection between forward homothetic processes, ergodic
control and ergodic BSDE was established. Other developments related to
multi-scale ill-posed HJB\ equations and to entropic risk measures can
be found in the aforementioned papers. We note that the discrete case developed in \cite{AngoshtariZariphopoulouZhouXX} and further studied in \cite{StrubZhouXX} is
particularly challenging as there is no infinitesimal stochastic calculus
and, furthermore, there are no general results for the functional equations
therein.

Despite the various technical difficulties, the concept of forward
performance criteria is well defined for stochastic optimization settings
whose classical (backward) analogues satisfy the DPP, and thus the
martingale/supermartingale properties as well as time-consistency hold.
However, these fundamentally interlinked connections break down when the
backward problems are time-inconsistent. 

Time-inconsistency is an important feature that arises in a plethora of
interesting problems in classical and behavioral finance. Among others, it is present in mean-variance optimization, hyperbolic discounting, and risk preferences involving probability distortions. Given, from the one hand, the recent developments in forward performance criteria and, from the other, the importance of time-inconsistent problems, an interesting question thus arises, namely, whether and how one can develop the concept of forward performance criteria for such settings. Herein, we study this question in the realm of \textit{rank-dependent} utilities.

Rank-dependent utility (RDU) theory was developed by \cite%
{Quiggin82,Quiggin93}, see also \cite{Schmeidler89}, and constitutes one of
the most important alternative theories of choice under risk to the expected
utility paradigm. It features two main components: a concave utility
function that ranks outcomes and a probability distortion function.
Rank-dependent utility theory is able to explain a number of empirical
phenomena such as the Allais paradox, the simultaneous investment in
well-diversified funds and poorly-diversified portfolios of stocks and low stock market participation (\cite{Polkovnichenko05}) and preference for skewness (\cite{BarberisHuang08}, \cite{DimmockKouwenbergMitchellPeijnenburgXX}).

Solving portfolio optimization problems under rank-dependent utility preferences is difficult because such problems are both \textit{time-inconsistent} and \textit{non-concave} due to the probability distortion. The difficulty of non-concavity was overcome by the quantile approach developed in \cite{JinZhou08}, \cite{CarlierDana11}, \cite{HeZhou11,HeZhou16} and \cite{Xu16}. A general solution for a rank-dependend utility maximization problem in a complete market was derived in \cite{XiaZhou16} and its effects on optimal investment decisions were extensively studied in \cite{HeKouwenbergZhou17,HeKouwenbergZhou18}. On the other hand, it remains an open problem to solve portfolio optimization problems under rank-dependent utility in general incomplete markets, where one can not apply the martingale approach and time-inconsistency thus becomes a real challenge.

The difficulties in developing forward rank-dependent criteria are both
conceptual and technical. Conceptually, it is not clear what could replace
the martingality/supermartingality requirements given that, in the classical
setting, the DPP fails. Furthermore, there is not even a notion of
(super)martingale under probability distortion. In addition, it is not clear how time-consistency could be incorporated, if at all. From the technical point of view, challenges arise due to the fact that probability distortions
are not amenable to infinitesimal stochastic calculus, which plays a key role in deriving the forward stochastic PDE.

We address these difficulties by first proposing two \textit{distinct}
definitions for a pair of processes, $\left( \left( u_{t}\left( x\right)
\right) _{t\geqslant 0},\left( w_{s,t}\left( p\right) \right) _{0\leqslant s<t}\right)
,$ $x\geqslant 0,$ $p\in \left[ 0,1\right] ,$ $t\geqslant 0,$ to be a forward
rank-dependent criterion. The first component, $u_{t}\left( x\right) ,$ is
the utility process while the second, $w_{s,t}\left( p\right) ,$ plays the
role of the probability distortion The first definition imitates the
martingale/supermartingale properties and requires, for all times, analogous
conditions but under the distorted conditional probabilities (cf.
Definition \ref{def:ForwardRDU}). It is based on the preservation of value along optimal
policies, and its loss along suboptimal ones. On the other hand, the second
definition is related to time-consistency. It uses a continuum of
optimization problems under the candidate pair $\left( u_{t}\left( x\right)
,w_{s,t}\left( p\right) \right) $ and requires time-consistency of the
candidate policies across any sub-horizon (cf.
Definition \ref{def:TimeConsistentRDPP}). 

We note that in both definitions, the utility process $u_{t}\left( x\right) $
is defined for all times $t\geqslant 0$ while the probability distortion $%
w_{s,t}\left( p\right) $ for all intermediate times $0\leqslant s<t.$ We also note that we consider deterministic processes $\left( u_{t}\left( x\right) ,w_{s,t}\left( p\right) \right) $ herein. This is important conceptually since it is not clear how to evaluate a stochastic utility function in the presence of probability distortion. Considering deterministic processes also allows us to develop a direct connection to deterministic, time-monotone forward criteria without probability distortion, as we show herein.

Naturally, in the absence of probability distortion, i.e., at the degenerate
case $w_{s,t}\left( p\right) \equiv p,$ $0\leqslant s<t,$ both definitions
reduce to the existing one of the forward performance criterion.
Surprisingly, it turns out that the two definitions we propose are also 
equivalent even in the non-degenerate case, as we show in
Proposition \ref{prop:TimeConsistency}. 

This is one of the first novel features in our analysis as it relates
probability distortions, which so far have been yielding time-inconsistency
of policies, to time-consistent optimal behavior. This unexpected connection
is also used for establishing new results for the so-called \textit{dynamic utilities} in the traditional (backward) setting, as we explain later on.

The second main result is the derivation of necessary conditions for a
distortion process $w_{s,t}$ to belong to a forward rank-dependent pair. We
establish a ``bifurcation" result, specifically, it is either the case that,
for \textit{all} $0\leqslant s<t,$ we must have 
\begin{align}
w_{s,t}(p)=\frac{1}{\mathbb{E}\left[ \rho _{s,t}^{1-\gamma }\right] }%
\int_{0}^{p}\left( \left( F_{s,t}^{\rho }\right) ^{-1}(q)\right) ^{1-\gamma
}dq,  \label{intro}
\end{align}%
for some $\gamma >0$ and with $\rho _{s,t}$ being the pricing kernel and $%
F_{s,t}^{\rho }$ its cumulative distribution, or it must be that the
inequality 
\begin{align}
w_{s,t}(p)\geqslant \mathbb{E}\left[ \rho _{s,t}\boldsymbol{1}_{\left \{ \rho
_{s,t}\text{ }\leqslant \text{ }\left( F_{s,t}^{\rho }\right) ^{-1}(p)\right \} }%
\right]   \label{intro-2}
\end{align}%
is always satisfied. These are the only two viable cases and they imply rather
distinct behavior with regards to the optimal allocation, with the latter family yielding zero allocation in the risky assets at all times. We will be referring to $\gamma $ as the \textit{(investor-specific) distortion parameter}.

For the market considered herein, the probability distortion process (\ref%
{intro-2}) reduces to \textit{\ }%
\begin{align}
w_{s,t}(p)=\Phi \left( \Phi ^{-1}\left( p\right) +\left( \gamma -1\right) 
\sqrt{\int_{s}^{t}\Vert \lambda _{r}\Vert ^{2}dr}\right) ,  \label{intro-3}
\end{align}%
with $\Phi $ being the cumulative normal distribution and $\lambda $ being
the market price of risk. This is a rather interesting formula as it
connects $w_{s,t}$ with the popular, in the traditional setting, Wang's
probability distortion (see \cite{Wang00}), which is of form $\Phi \left(
\Phi ^{-1}\left( .\right) +a\right) ,$ for some fixed $a.$ 

In the forward setting, however, $w_{s,t}$ is affected not only by the
distortion parameter $\gamma ,$ which is chosen by the investor, but also by
the current market behavior, as manifested by the \textit{market-specific input} $%
A_{s,t}:=$ $\sqrt{\int_{s}^{t}\Vert \lambda _{r}\Vert ^{2}dr}.$ This is
intuitively pleasing, for forward criteria are expected to follow the market
in ``real-time". Furthermore, the multiplicative coefficient $\left( \gamma
-1\right) $ combines in a very transparent way the market condition with the
investor's attitude. The latter can be thought as \textit{objective} $\left( \gamma
=1\right) ,$ \textit{pessimistic} $\left( \gamma <1\right) $ or \textit{optimistic} $\left(
\gamma >1\right).$

As a corollary to the above results, we obtain that the distortion process of any forward rank-dependent  satisfies the monotonicity condition of \cite{JinZhou08}, cf. Assumption 4.1 and the discussion in Section 6.2 therein. This implies in particular that the optimal wealth process is strictly decreasing as a function of the pricing kernel. An interesting analogy is a result of \cite{XiaZhou16} showing that the Jin-Zhou monotonicity condition is also automatically satisfied for a representative agent of an Arrow-Debreu economy. In other words, we have that the monotonicity condition of \cite{JinZhou08} is satisfied if the market is exogenously given and the preferences are endogenously determined through the framework of forward criteria, or if the preferences are exogenously given and the pricing kernel is endogenously determined through an equilibrium condition.

The third main result is the actual \textit{construction} of forward rank-dependent criteria. In the degenerate case (\ref{intro-3}), it follows easily that $u_{t}\left( x\right) =u_{0}\left( x\right) ,$ $t\geqslant 0,$ for  the optimal investment in all risky assets is always zero. In the non-degenerate case (\ref{intro-2}),
we establish a direct equivalence with deterministic, time-monotone forward criteria in the
absence of probability distortions. Specifically, for a given $\gamma ,$ we
introduce a new measure, the $\gamma $-\textit{distorted measure} and a
related \textit{distorted market }with modified risk premium $\tilde{\lambda}%
_{t}:=\gamma \lambda _{t}$ (see Subsection \ref{subsec:AuxiliaryMarket}). As we explain later on, the
motivation for considering these measure and market variations comes from the
form of the optimal wealth process for the non-degenerate case.

In the distorted market, we in turn recall the standard (no probability
distortion) time-monotone forward criterion, denoted by $U_{t}(x).$ As
established in \cite{MusielaZariphopoulou10a}, it is given by $U_{t}(x)=v(x,\int_{s}^{t}\Vert 
\tilde{\lambda}_{r}\Vert ^{2}dr),$ with the function $v\left( x,t\right) $
satisfying $v_{t}=\frac{1}{2}\frac{v_{x}^2}{v_{xx}}.$ Herein, we establish
that 
\begin{align}
u_{t}\left( x\right) =U_{t}(x)=v \left(x,\int_{s}^{t}\gamma ^{2}\Vert \lambda
_{r}\Vert ^{2}dr \right).  \label{intro-4}
\end{align}%
In other words, the utility process $u_{t}\left( x\right) $ of the forward
rank-dependent criterion in the original market corresponds to a
deterministic, time-monotone forward criterion in a pseudo-market with modified risk premia and vice-versa. 

If the investor is objective $\left( \gamma =1\right),$ there is no
probability distortion and, as a result, the two markets become identical
and the two criteria coincide, $u_{t}\left( x\right) =U_{t}(x).$ For
optimistic investors $\left( \gamma >1\right) ,$ however, the
time-monotonicity of the function $v(x,t)$ results in a more pronounced
effect on how the forward rank-dependent utility decays with time.
Specifically, the higher the optimism (higher $\gamma ),$ the larger the
time-decay in the utility criterion, reflecting a higher loss of
subjectively viewed better opportunities. The opposite behavior is observed
for pessimistic investors $\left( \gamma <1\right) $ where the time-decay is
slower, since the market opportunities look subjectively worse. Finally, the
limiting case $\gamma =0$ corresponds to a subjectively worthless distorted\
market. The latter yields $U_{t}(x)=U_{0}\left( x\right) =v\left( x,0\right)
,$ and in turn (\ref{intro-4}) implies that $u_{t}\left( x\right)
=u_{0}\left( x\right) ,$ $t\geqslant 0.$ 

In addition to the construction approach, the equivalence established in Theorem \ref{thm:ClassicalForwardUtilityEquivalentMeasure} yields explicit formulae for the optimal wealth and investment policies under forward probability distortions by using the analogous formulae under deterministic, time-monotone forward criteria. 

As mentioned earlier, our construction of time-consistent
criteria even in the presence of probability distortion prompts us to
revisit the classical (backward) setting and investigate if and how our
findings can be used to build time-consistent policies therein. The  \textit{dynamic utility approach} developed in \cite{KarnamMaZhang17} seems suitable to this end. This approach builds on the observation that the time-inconsistency of stochastic optimization problems is partially due to the following restriction: The utility functional determining the objective at an intermediate time is essentially the same as the utility functional at initial time modulus conditioning on the filtration. The dynamic utility approach relaxes this restriction and allows the intermediate utility functional to vary more freely so that the DPP holds. In a recent work closely related to this paper, \cite{MaWongZhangXX} introduce a \textit{dynamic distortion function}. This leads to a distorted conditional expectation which is time-consistent in the sense that the tower-property holds. In their setting, an It\^{o} process is given and fixed and the dynamic distortion function is then constructed for this particular process. Since the construction depends on the drift and volatility parameters of the It\^{o} process, their results are not directly applicable to our setting, where we consider an investment problem and the state process is not a priori given, but instead controlled by the investment policy. Herein, we extend the construction of dynamic distortion functions to controlled processes for the problem of rank-dependent utility maximization in a financial market with determinstic coefficients. We find that constructing a dynamic utility which is restricted to remain in the class of RDU preference functionals is possible if and only if the initial probability distortion function belongs to the family introduced in \cite{Wang00}. 

Studying time-inconsistency induced by distorting probabilities is one of the remaining open challenges for the psychology of tail events identified in the review article \cite{Barberis13}. We contribute to this research direction by developing a new class of risk preferences and showing that investment under probability can be time-consistent. Furthermore, we fully characterize the conditions under which this is possible, namely if and only if the marginal probability distortion equals to a normalized power of the quantile function of the pricing kernel.

The paper is organized as follows. In Section \ref{sec:ModelAndFRDPP}, we describe the model and review the main results for the classical rank-dependent utility. In Section \ref{sec:ForwardRDU}, we introduce the definitions of the forward rank-dependent performance criteria and establish their equivalence. We continue in Section \ref{sec:CharacterizationDistortionFunction}, where we derive the necessary conditions for a distortion probability process to belong to a forward rank-dependent pair. In Section \ref{sec:ForwardRDUDistortedMeasure}, we establish the connection with the deterministic, time-monotone forward criteria, the form of the optimal wealth and portfolio processes and provide examples. In Section \ref{sec:DynamicUtilityApproach}, we relate
our results to dynamic utility approach and show that constructing a dynamic utility restricted to remain within the class of RDU preference functionals is possible if and only if the initial distortion function belongs to the class introduced in \cite{Wang00}. We conclude in Section \ref{sec:Conclusions}. To ease the presentation, we delegate all proofs in an Appendix.

\section{The investment model and background results}

\label{sec:ModelAndFRDPP}We start with the description of the market model and a review of the main concepts and results for rank-dependent utilities.

The financial market consists of one risk-free and $N$ risky assets. The
price of the $i^{th}$ risky asset solves 
\begin{align}
dS_{t}^{i}=S_{t}^{i}\left( \mu _{t}^{i}dt+\sum_{j=1}^{N}\sigma
_{t}^{ij}dW_{t}^{j}\right) ,\quad t\geqslant 0,  \label{stock}
\end{align}%
with $S_{0}^{i}=s_{0}^{i}>0$, $i=1,\dots ,N$. The process $W=(W_{t})_{t\geqslant
0}$ is an $N$-dimensional Brownian motion on a filtered probability space $\left(
\Omega ,\mathcal{F},\mathbb{F},\mathbb{P}\right) $ satisfying the usual conditions and where $\mathbb{F}=(\mathcal{F}%
_{t})_{t\geqslant 0}$ is the completed filtration generated by $W$.

The drift and volatility coefficients are assumed to be deterministic
functions that satisfy $\int_{0}^{t}|\mu _{s}^{i}|ds<\infty $ and $%
\int_{0}^{t}\left( \sigma _{s}^{ij}\right) ^{2}ds<\infty $, $t\geqslant 0$ and $%
i,j=1,\dots ,N$. We denote the volatility matrix by $\sigma _{t}:=\left(
\sigma _{t}^{ij}\right) _{N\times N}$. We assume that $\sigma _{t}$ is
invertible for all $t\geqslant 0,$ to ensure that the market is arbitrage free
and complete. We also define the \textit{market price of risk} process, 
\begin{align}
\lambda _{t}:=\sigma _{t}^{-1}\mu _{t}  \label{lambda}
\end{align}%
and assume that $\lambda _{t}>0,$ $t\geqslant 0$.

For each $t>0$, we consider the (unique) pricing kernel 
\begin{align}
\rho _{t}=\exp \left( -\int_{0}^{t}\frac{1}{2}\Vert \lambda _{r}\Vert
^{2}dr-\int_{0}^{T}\lambda _{r}^{\prime }dW_{r}\right) .
\label{kernel-single}
\end{align}%
For $0<s\leqslant t$, we further define 
\begin{align}
\rho _{s,t}:=\frac{\rho _{t}}{\rho _{s}}=\exp \left( -\int_{s}^{t}\frac{1}{2}%
\Vert \lambda _{r}\Vert ^{2}dr-\int_{s}^{t}\lambda _{r}^{\prime
}dW_{r}\right) .  \label{kernel-dynamic}
\end{align}%
We also denote the cumulative distribution function of $\rho _{t}$ and $\rho
_{s,t}$ by $F_{t}^{\rho }$ and $F_{s,t}^{\rho }$ respectively. \ 

We stress, and this will be also discussed later on, that while we
pre-assume that the market coefficients are deterministic processes, we do 
\textit{not }pre-specify their values. This is in contrast with the
classical setting where the full model (or a plausible family of models) needs to be determined at initial time and for the entire trading horizon,
and thus the exact dynamics of $\mu _{t}^{i}$ and $\sigma _{t}^{ij},$ $%
i,j=1,\dots ,N,$ have to be a priori known.

The agent starts at $t=0$ and trades between the riskless and the risky
assets, using a self-financing trading policy $(\pi _{t}^{0},\pi
_{t})_{t\geqslant 0},$ where $\pi _{t}^{0} = \pi_{t}^{0} (\omega; x)$ denotes the allocation in the
riskless asset and the vector $\pi _{t}:=\left( \pi _{t}^{1},\pi
_{t}^{2},...,\pi _{t}^{N}\right) $, with $\pi _{t}^{i} = \pi^{i}_t (\omega;x),$ $i=1,\dots ,N,$
representing the amount invested, at time $t,$ in the risky asset $i.$ 
Strategies are allowed to depend on the initial wealth $x > 0$ and the state of the world $\omega \in \Omega$. We usually drop the $\omega$ and $x$ argument whenever the context is clear.  In turn, the wealth process $X=(X_{t}^{x,\pi })_{t\geqslant 0}$ solves the
stochastic differential equation 
\begin{align}
dX_{t}^{x,\pi }=\pi _{t}^{\prime }\mu _{t}dt+\pi _{t}^{\prime }\sigma
_{t}dW_{t},\quad t\geqslant 0,  \label{wealth-SDE}
\end{align}%
with $X_{0}^{x,\pi }=x.$ For notational simplicity, we will often write $%
X^{\pi }$ instead of $X^{x,\pi }.$

The set of admissible strategies is defined as 
\begin{align}\label{admissible-set}
\begin{split}
\mathcal{A} := \bigg\{  \pi & \big\vert \pi _{t}\text{ is }\mathbb{F}\text{%
-progressively measurable,}\\
& \qquad \text{with}\int_{0}^{t}\Vert \pi _{s}\Vert ^{2}ds<\infty 
\text{ and }X_{t}^{x,\pi }\geqslant 0,\text{ for }t\geqslant 0,\text{ }x\geqslant 0\text{ }%
\bigg\} .
\end{split}
\end{align}%
For a given time, $t_{0},$ and an admissible policy, $\widetilde{\pi},$ we
also introduce 
\begin{align}
\mathcal{A}(\widetilde{\pi },t_{0}):=\{ \pi \in \mathcal{A}|\pi _{s}\equiv 
\widetilde{\pi }_{s},s\in \lbrack 0,t_{0}]\},
\label{admissible-set-truncated}
\end{align}%
namely, the set of admissible strategies which \textit{coincide} with this specific policy in $\left[ 0,t_{0}\right] $.

\subsection{Rank-dependent utility theory}

To ease the presentation and build motivation for the upcoming analysis, we
start with a brief overview of the rank-dependent utilities and the main
results on portfolio optimization under such preferences for the market
considered herein.

The rank-dependent utility value of a prospect $X$ is defined as 
\begin{align}
V(X):=\int_{0}^{\infty }u(\xi )d\left( -w\left( 1-F_{X}(\xi )\right) \right)
,  \label{eq:RDU-PreferenceValue}
\end{align}
where $u$ is a \textit{utility function} and $w$ is a \textit{probability
distortion function}, cf. \cite{Quiggin82,Quiggin93} and \cite{Schmeidler89}.

We assume that $u$ and $w$ belong to the sets $\mathcal{U}$ and $\mathcal{W}%
, $ introduced next.

\begin{definition}\label{def:UtilityAndDistortionFunctions} Let $\mathcal{U}$ be
the set of all utility functions $u:\left[ 0,\infty \right) \rightarrow 
\mathbb{R},$ with $u$ being strictly increasing, strictly concave, twice
continuously differentiable in $\left( 0,\infty \right) ,$ and satisfying
the Inada conditions $\lim_{x\downarrow 0}u^{\prime }(x)=\infty $ and $%
\lim_{x\uparrow \infty }u^{\prime }(x)=0$.

Let $\mathcal{W}$ be the set of probability distortion functions $%
w:[0,1]\rightarrow \lbrack 0,1]$ that are continuously differentiable,
strictly increasing and satisfying $w(0)=0$ and $w(1)=1$.
\end{definition}

At initial time $t=0$, an agent chooses her investment horizon $T>0,$  the
dynamics in (\ref{stock}) for $\left[ 0,T\right]$, together with $u\in 
\mathcal{U}$ and $w\in \mathcal{W}.$ She then solves the portfolio
optimization problem 
\begin{align}
v(x,0)=\sup_{\pi \in \mathcal{A}_{T}}V(X_{T}^{\pi })  \label{prob:RDU-backward}
\end{align}%
with $X_{s}^{\pi },s\in \left[ 0,T\right] ,$ solving (\ref{wealth-SDE}) and $%
X_{0}^{\pi }=x,$ $V$ is given in (\ref{eq:RDU-PreferenceValue}), and $%
\mathcal{A}_{T}$ is defined similarly to $\mathcal{A}$ in (\ref%
{admissible-set}), up to horizon $T$.

This problem has been studied by various authors; see, among others, \cite{CarlierDana11}, \cite{XiaZhou16}, \cite{Xu16} or \cite%
{HeKouwenbergZhou17,HeKouwenbergZhou18}. Fundamental
difficulties arise from the time-inconsistency which stems from the probability distortion. Consequently, key elements in stochastic optimization, like the Dynamic Programming Principle, the Hamilton-Jacobi-Bellman (HJB)\ equation, the martingality of the value function process along an optimum process and others, are lost.

The analysis of (\ref{prob:RDU-backward}) has been carried out using a well known reformulation to a static problem and the quantile method developed in \cite{JinZhou08}, \cite{CarlierDana11}, \cite{HeZhou11,HeZhou16} and \cite{Xu16}. Specifically, because the market is complete,
any $\mathcal{F}_{T}$-measurable prospect $X$ that satisfies the \textit{%
budget constraint} $\mathbb{E}\left[ \rho _{T}X\right] =x$ can be
replicated by a self-financing policy. In turn, problem (\ref%
{prob:RDU-backward}) reduces to%
\begin{align}
\sup_{X}V\left( X\right) \text{ \  \ with \  \ }\mathbb{E}\left( \rho
_{T}X\right) \leqslant x,\text{ }X\geqslant 0,\text{ }X\in \mathcal{F}_{T}.
\label{RDU-static}
\end{align}%
One of the main steps in its solution are the specification of the ``optimal"
Lagrange multiplier, the construction of the terminal optimal wealth and the
characterization of the optimal policy through martingale representation
results. The rank-dependent case, however, is considerably harder due, from the one hand, the
joint nonlinearities (risk preferences and non-linear averaging) in
criterion (\ref{eq:RDU-PreferenceValue}) and, from the other, the non-concavity due to the presence of the probability distortion.

A very important feature in the RDU\ family of preferences is that, for
certain choices of the probability distortion function $w$, the optimal
investment in the risky assets turns out to be zero, even if the market
price of risk is not zero. The optimal wealth then remains unchanged (recall
that interest rate is taken to be zero). We will be referring to this as a
``degenerate optimal investment case". Note that this is in direct contrast
with the classical setting where a risk averse agent would always invest in
a worthy (non-zero risk premium) market.

Central results on the optimal investment case were derived
in \cite{XiaZhou16} and \cite{Xu16} (see also \cite{CarlierDana11}) and are stated next.

\begin{theorem}
\label{thm:BackwardRDEUSolution} Let $u\in \mathcal{U}$ and $w\in \mathcal{W}
$. If there exists an optimal wealth to (\ref%
{RDU-static}), it is given by 
\begin{align}
X_{T}^{\ast }= \left(u^{\prime} \right)^{ -1}\left( \lambda ^{\ast }\hat{N}^{\prime }\left(
1-w\left( F_{T}^{\rho }(\rho _{T})\right) \right) \right) ,
\label{optimalX-backward}
\end{align}%
where $\hat{N}$ is the concave envelope of 
\begin{align}
N(z):=-\int_{0}^{w^{-1}(1-z)}(F_{T}^{\rho })^{-1}(t)dt,\quad z\in \lbrack
0,1].  \label{envelope-backward}
\end{align}%
and the Lagrangian multiplier $\lambda ^{\ast }>0$ is determined by $\mathbb{%
E}[\rho _{T}X_{T}^{\ast }]=x$.
\end{theorem}

\bigskip

We also recall that if, in addition, the so called \textit{Jin-Zhou
monotonicity condition }holds, namely, if the function $f:\left[ 0,1\right]
\rightarrow \mathbb{R}^{+},$ defined as 
\begin{align}
f\left( p\right) :=\frac{(F_{T}^{\rho })^{-1}(p)}{w^{\prime }(p)}
\label{Jin-Zhou}
\end{align}%
is nondecreasing, then equality (\ref{optimalX-backward}) simplifies to 
\begin{align}
X_{T}^{\ast }=\left( u^{\prime }\right)^{-1}\left( \lambda ^{\ast }\frac{%
\rho _{T}}{w^{\prime }\left( F_{T}^{\rho }(\rho _{T})\right) }\right) ,
\label{alth-backward-simplified}
\end{align}%
see Remark 3.4 in \cite{XiaZhou16} or \cite{JinZhou08}.

Further results for problem (\ref{RDU-static}) were derived in \cite%
{XiaZhou16}, where it was shown that if, for each $\lambda >0,$ the
inequality 
\begin{align*}
\mathbb{E}\left[ \rho _{T}\left( u^{\prime }\right) ^{-1}\left( \lambda \hat{%
N}^{\prime }\left( 1-w\left( F_{T}^{\rho }(\rho _{T})\right) \right) \right) %
\right] <\infty
\end{align*}%
holds, with $\hat{N}^{\prime }$ as in (\ref{envelope-backward}), then an
optimal solution exists and is of form (\ref{optimalX-backward}).

In addition, the author in \cite{Xu16} showed that the existence of a
non-degenerate optimal investment policy is \textit{equivalent} to the
existence of a Lagrangian multiplier $\lambda ^{\ast }$ and that, in this
case, the terminal optimal wealth is as given in Theorem \ref%
{thm:BackwardRDEUSolution}.

\section{Forward rank-dependent performance criteria}

\label{sec:ForwardRDU} We introduce the concept of forward performance
criteria in the framework of rank-dependent preferences. We first review the
definition of the forward performance criterion (slightly modified for the
setting and notation herein); see, among others, \cite{MusielaZariphopoulou06,MusielaZariphopoulou09,
MusielaZariphopoulou10a}. We then
discuss the various difficulties in extending this concept when probability
distortions are incorporated.

\begin{definition}
\label{def:ForwardEUT} An $\mathbb{F}$-adapted process $(U_{t})_{t \geqslant 0}$ is a forward performance criterion if

\begin{enumerate}
\item[i)] for any $t\geqslant 0$ and fixed $\omega \in \Omega $, $U_{t}\in 
\mathcal{U},$

\item[ii)] for any $\pi \in \mathcal{A}$,  $0 \leqslant s \leqslant t$ and $x > 0$
\begin{align}
\mathbb{E}\left[ \left. U_{t}\left( X_{t}^{x,\pi }\right) \right \vert 
\mathcal{F}_{s}\right] \leqslant U_{s}\left( X_{s}^{x,\pi }\right) ,
\label{supermartingality}
\end{align}

\item[iii)] there exists $\pi ^{\ast }\in \mathcal{A}$ such that, for any $%
0\leqslant s\leqslant t,$ and $x > 0,$
\begin{align}
\mathbb{E}\left[ \left. U_{t}\left( X_{t}^{x,\pi ^{\ast }}\right)
\right \vert \mathcal{F}_{s}\right] =U_{s}\left( X_{s}^{x,\pi ^{\ast
}}\right) .  \label{martingality}
\end{align}
\end{enumerate}
\end{definition}

The above definition was directly motivated by the DPP, a key feature in the
classical stochastic optimization, which yields the above supermartingality
and martingality properties of the value function process along an
admissible and an optimal policy, respectively. Furthermore, directly
embedded in this fundamental connection between DPP\ and (\ref%
{supermartingality}) and (\ref{martingality}), is the time-consistency of
the optimal policies.

Once, however, probability distortions are incorporated, none of these
features exist in the classical rank-dependent case, as we discussed in the
previous section. Indeed, the DPP\ does not hold and, naturally,
time-inconsistency arises. Furthermore, there is no general notion of
supermartingality and martingality under probability distortions, and thus
it is not clear what the analogues of (\ref{supermartingality}) and (\ref%
{martingality}) are\footnote{%
In a recent work by \cite{MaWongZhangXX} the concept of nonlinear expectation and time-consistency was studied in a specific setting. We refer to these results in Section \ref{sec:DynamicUtilityApproach} herein, where we also provide some new results in this
direction.}. \ In other words, we lack the deep connection among the DPP, the martingality/supermartingality of the value function process, and the time-consistency of the optimal policies, which is the cornerstone in the expected utility paradigm. Thus, it is not at all clear how to define the forward rank-dependent performance criteria. We address these difficulties in two steps.

We first propose a definition of forward rank-dependent criteria by directly imitating requirements (\ref{supermartingality}) and (\ref{martingality}).
Specifically, we propose (\ref{supermartingality-RD}) and (\ref%
{martingality-RD}), respectively, where we use (conditional) distorted
probabilities instead of the regular ones. This definition is a natural, direct analogue to Definition \ref{def:ForwardEUT}, as it is built on the preservation of value along an optimal policy and its decay along a suboptimal one. The novel element in Definition \ref{def:ForwardRDU} is that we seek a \textit{%
pair} of processes $\left( \left( u_{t}\right) _{t\geqslant 0},\left(
w_{s,t}\right) _{0\leqslant s<t}\right),$ corresponding to a dynamic ``forward utility" $u_{t}$ that is defined for \textit{each time}, say $t\geqslant 0,$ and
a dynamic ``forward probability distortion" $w_{s,t}$ that is defined for 
\textit{all intermediate }times $s\in \left[ 0,t\right) .$ In other words,
while $u_{t}$ is parametrized solely by $t,$ the second component $w_{s,t}$
is parametrized by \textit{both} the starting and the end points, $s$ and $t.
$

Note that this definition is not superfluous. Indeed, if we choose $%
w_{s,t}(p)\equiv p,$ for all $t\geqslant 0$ and $0\leqslant s<t,$ then Definition \ref{def:ForwardRDU} reduces to Definition \ref{def:ForwardEUT} above; see Proposition \ref{prop:DefinitionForwardEquivalent} below.

Definition \ref{def:ForwardRDU}, however, does not give any insights about the time-consistency of the optimal policies. As a matter of fact, it is not even clear whether we should even seek such a property given that, after all, time-consistency does not hold in the classical rank-dependent setting.

Surprisingly, it turns out that we can actually build a direct connection between time-consistency and forward rank-dependent criteria. For this, we first introduce the concept of \textit{time-consistent }rank-dependent processes and, subsequently, a subclass of this family that preserve the forward performance value along an optimal policy; see parts (i) and (ii) in Definition \ref{def:TimeConsistentRDPP}, respectively.

In turn, we show in Proposition \ref{prop:TimeConsistency} that Definitions \ref{def:ForwardRDU} and \ref{def:TimeConsistentRDPP} are equivalent. In other words, we establish an equivalence between forward rank-dependent performance criteria and the time-consistent ones that also preserve the performance value.

Finally, we note that herein we work exclusively with \textit{deterministic}
processes for both $u_{t}$ and $w_{s,t}.$ We do this for various reasons.
Firstly, it is assumed that the coefficients of the risky assets are
deterministic (cf. (\ref{stock})) and, therefore, it is natural to first explore the class of deterministic forward rank-dependent criteria.
Secondly, working with deterministic criteria enables us to build a direct connection with time-monotone analogues that are also deterministic.
Thirdly, it is not yet clear how to define non-deterministic criteria even for the market herein. We recall that in the standard forward case, stochasticity arises both from the market dynamics and the forward volatility process, which is an investor-specific input. It is conceptually unclear how to evaluate a stochastic preference functional when probability distortions are incorporated.

Next, we introduce some notation and provide the relevant definitions and results. To this end, we denote by $\mathbb{P}\left[ \left. \cdot
\right \vert \mathcal{G}\right] (\omega )$ the  conditional probability given
a sigma-algebra $\mathcal{G}\subseteq \mathcal{F}.$ For a random variable $%
X\in \mathcal{F}$, we denote by $F_{X|\mathcal{G}}(\cdot ;\omega )=\mathbb{P}%
\left[ X\leqslant \cdot |\mathcal{G}\right] (\omega )$ the regular conditional
distribution function of $X$ given $\mathcal{G}$\footnote{
We will also use the standard convention that $\int_{0}^{\infty }u(\xi )d\left( -w\left( 1-F_{X|\mathcal{G}}(\xi )\right)
\right) =-\infty $
whenever $
\int_{0}^{\infty }\max \left( 0,-u(\xi )\right) d\left( -w\left( 1-F_{X|%
\mathcal{G}}(\xi )\right) \right) =\int_{0}^{\infty }\max \left( 0,u(\xi
)\right) d\left( -w\left( 1-F_{X|\mathcal{G}}(\xi )\right) \right) =\infty ,
$
for $u\in \mathcal{U}$ and $w\in \mathcal{W}.$}.

\begin{definition}
\label{def:ForwardRDU} A pair of deterministic
processes $\left( \left( u_{t}\right) _{t\geqslant 0},\left( w_{s,t}\right)
_{0\leqslant s<t}\right) $ is a forward rank-dependent performance criterion if the following properties hold:

\begin{enumerate}
\item[i)] for any $t\geqslant 0$, $u_{t}(\cdot )\in \mathcal{U}$ and for any $0 \leqslant s < t$ $w_{s,t}(\cdot )\in \mathcal{W}$%
.

\item[ii)] for any $\pi \mathcal{\in A},$ $0\leqslant s<t$ and $x>0,$
\begin{align}
\int_{0}^{\infty }u_{t}(\xi )d\left( -w_{s,t}\left( 1-F_{X_{t}^{x,\pi }|%
\mathcal{F}_{s}}(\xi )\right) \right) \leqslant u_{s}\left( X_{s}^{x,\pi }\right)
.  \label{supermartingality-RD}
\end{align}

\item[iii)] there exists $\pi ^{\ast }\mathcal{\in A}$, such that for any $0\leqslant s<t$ and $x>0,$
\begin{align}
\int_{0}^{\infty }u_{t}(\xi )d\left( -w_{s,t}\left( 1-F_{X_{t}^{x,\pi ^{\ast
}}|\mathcal{F}_{s}}(\xi )\right) \right) =u_{s}\left( X_{s}^{x,\pi ^{\ast
}}\right) .  \label{martingality-RD}
\end{align}
\end{enumerate}
\end{definition}

\bigskip

Naturally, if there is no probability distortion, the above definition
should reduce to Definition \ref{def:ForwardEUT}.

\begin{proposition}
\label{prop:DefinitionForwardEquivalent} i) Let $\left( \left( u_{t}\right)
_{t\geqslant 0},\left( w_{s,t}\right) _{0\leqslant s<t}\right) $ be a forward
rank-dependent performance criterion with $w_{s,t}(p)\equiv p,$ $p\in \lbrack
0,1]$ and $0\leqslant s<t$. Then $\left( u_{t}\right) _{t\geqslant 0}$ is a forward
performance process.

ii)\ Conversely, if $\left( u_{t}\right) _{t\geqslant 0}$ is a deterministic
forward performance process, then the pair $\left( \left( u_{t}\right)
_{t\geqslant 0},\left( w_{s,t}\right) _{0\leqslant s<t}\right) ,$ with $w_{s,t}(p)\equiv p$, $p\in \lbrack 0,1]$ for all $0\leqslant s<t$, is a forward
rank-dependent performance criterion.
\end{proposition}

\textbf{\ }

We now present an alternative definition.

\begin{definition}\label{def:TimeConsistentRDPP}
Let $\left( u_{t}\right) _{t\geqslant 0}$ and $\left( w_{s,t}\right) _{0\leqslant s<t}$
be deterministic processes with $u_{t}\in \mathcal{U}$ and $w_{s,t}\in 
\mathcal{W}.$ 

\begin{enumerate}
\item[i)] A pair $\left( \left( u_{t}\right) _{t\geqslant 0},\left(
w_{s,t}\right) _{0\leqslant s<t}\right) $ is called a time-consistent
rank-dependent performance criterion, if there exists $\pi ^{\ast }\in 
\mathcal{A}$, such that $\pi ^{\ast }$ solves the optimization problem 
\begin{align}
\max_{\pi \in \mathcal{A}(\pi ^{\ast },s)}\int_{0}^{\infty }u_{t}(\xi
)d\left( -w_{s,t}\left( 1-F_{X_{t}^{x,\pi }|\mathcal{F}_{s}}(\xi )\right)
\right)   \label{prob:TC-general}
\end{align}%
with $dX_{r}^{x,\pi }=\pi _{r}^{\prime }\mu _{r}dr+\pi _{r}^{\prime }\sigma
_{r}dW_{r},$ $r\in \lbrack 0,t]$ and $X_{0}^{x,\pi }=x,$ for any $%
0\leqslant s<t$ and $x > 0.$

\item[ii)] A pair $\left( \left( u_{t}\right) _{t\geqslant 0},\left(
w_{s,t}\right) _{0\leqslant s<t}\right) $ satisfying (i) is called a
time-consistent rank-dependent performance criterion preserving the
performance value if, for any optimal policy $\pi ^{\ast }$ as in (i), we
have 
\begin{align}
\int_{0}^{\infty }u_{t}(\xi )d\left( -w_{s,t}\left( 1-F_{X_{t}^{x,\pi ^{\ast
}}|\mathcal{F}_{s}}(\xi )\right) \right) =u_{s}\left( X_{s}^{x,\pi ^{\ast
}}\right) ,  \label{TC-forward}
\end{align}%
for any $0\leqslant s<t,$ and $x>0.$
\end{enumerate}
\end{definition}

\bigskip

Definition \ref{def:TimeConsistentRDPP} above is built around the time-consistency of optimal
policies (assuming that at least one such policy exists). It is also in
direct alignment with the DPP as we require that, for any investment horizon 
$t\geqslant 0$ and intermediate time $s\in \lbrack 0,t]$, the investment policy 
$\pi ^{\ast }=\left( \pi _{r}^{\ast }\right) _{0\leqslant r\leqslant t}$ optimizing
the rank-dependent utility value determined by $\left( u_{t}\mathbf{,}%
w_{0,t}\right) $, remains optimal over the investment interval $[s,t]$ with
respect to the rank-dependent utility described by $u_{t}$ and $w_{s,t}$.

The following Proposition states that Definitions \ref{def:ForwardRDU} and \ref{def:TimeConsistentRDPP} are actually
equivalent. 

\begin{proposition}
\label{prop:TimeConsistency} A pair of deterministic functions $\left(
\left( u_{t}\right) _{t\geqslant 0},\left( w_{s,t}\right) _{0\leqslant s<t}\right) $
is a forward rank-dependent performance criterion if and only if it is a
time-consistent rank-dependent performance criterion preserving the
performance value.
\end{proposition}

In the absence of probability distortion, properties (\ref{supermartingality}%
) and (\ref{martingality}), together with stochastic calculus yield a
stochastic PDE that the forward performance process $U_{t}\left( x\right) $
is expected to satisfy. This SPDE\ plays the role of the
Hamilton-Jacobi-Bellman (HJB)\ equation that arises in the backward setting.
In the rank-dependent case, however, these concepts and tools do not exist,
which makes the analysis considerably harder. Besides, we need to
characterize a pair of processes, the utility and the probability
distortion, and not just one. 

The methodology developed herein starts with a complete characterization of
all viable probability distortion functions.

\section{Characterization of the forward probability distortion functions }

\label{sec:CharacterizationDistortionFunction}

We present one of the main results herein, deriving \textit{necessary
conditions} for a deterministic probability distortion process, $w_{s,t},$
to belong to a forward rank-dependent pair $\left( \left( u_{t}\right)
_{t\geqslant 0},\left( w_{s,t}\right) _{0\leqslant s<t}\right) $. We first state all
pertinent results and then provide a discussion at the end of this section.

\begin{theorem}
\label{thm:WeightingNecessary} Let the pair $\left( \left( u_{t}\right)
_{t\geqslant 0},\left( w_{s,t}\right) _{0\leqslant s<t}\right) $ be a deterministic
forward rank-dependent performance criterion. Then, it is either the case that

i) there exists $\gamma >0$ such that, for each $p\in \left[ 0,1\right] ,$
and each $t\geqslant 0$ and $0\leqslant s<t,$ $w_{s,t}(p)$ is given by 
\begin{align}\label{distortion-forward}
w_{s,t}(p)=\frac{1}{\mathbb{E}\left[ \rho _{s,t}^{1-\gamma }\right] }%
\int_{0}^{p}\left( \left( F_{s,t}^{\rho }\right) ^{-1}(q)\right) ^{1-\gamma
}dq,  
\end{align}

or

ii) for each $p\in \left[ 0,1\right] ,$ and each $t\geqslant 0$ and $0\leqslant
s<t$, $w_{s,t}(p)$ satisfies 
\begin{align}
w_{s,t}(p)\geqslant \mathbb{E}\left[ \rho _{s,t}\boldsymbol{1}_{\left \{ \rho
_{s,t}\text{ }\leqslant \text{ }\left( F_{s,t}^{\rho }\right) ^{-1}(p)\right \} }%
\right] .  \label{distortion-forward-degenerate}
\end{align}
\end{theorem}

Unless it is stated otherwise, we assume for the remainder of the paper that
both processes $u_{t},w_{s,t}$ are deterministic.

The following result follows directly from (\ref{kernel-dynamic}).

\begin{corollary}\label{cor:Wang}
If the forward probability distortion is given
by (\ref{distortion-forward}), then %
\begin{align}
w_{s,t}(p)=\Phi \left( \Phi ^{-1}\left( p\right) +\left( \gamma -1\right) 
\sqrt{\int_{s}^{t}\Vert \lambda _{r}\Vert ^{2}dr}\right) ,
\label{Wang-dynamic}
\end{align}%
where $\left( \lambda _{t}\right) _{t\geqslant 0}$ is the
market price of risk (cf. (\ref{lambda})).
\end{corollary}

The following result yields that it is necessary to allow the family of
probability distortion functions of a forward rank-dependent performance
process to depend on both the initial and terminal time. Otherwise, forward
rank-dependent criteria reduce to the case without any probability
distortion. 

\begin{corollary}
\label{cor:WeightingIndependentInitialTime} Let $\left( \left(
u_{t}\right) _{0\leqslant t},\left( w_{s,t}\right) _{0\leqslant s<t}\right) $ be a
forward rank-dependent performance criterion such that, for each $t\geqslant 0,$ $%
w_{s,t}(p)=w_{r,t}(p)$ for all $0\leqslant s,r\leqslant t$ and $p\in \lbrack 0,1]$.
Then, it must be that $w_{s,t}(p)=p,$ for all $t\geqslant 0,$ $0\leqslant s<t,$ and $%
p\in \lbrack 0,1]$.
\end{corollary}

The following result yields the optimal wealth processes under the two cases
(\ref{distortion-forward}) and (\ref{distortion-forward-degenerate}), and an
admissible utility function $u_{t}.$

\begin{proposition}\label{prop:OptimalWealthLagrangianMultiplier}
Let the pair $\left( \left( u_{t}\right) _{t\geqslant 0},\left( w_{s,t}\right)_{0\leqslant s<t}\right) $ be a forward rank-dependent performance criterion. The following cases hold:

i) if $w_{s,t}$ satisfies (\ref{distortion-forward}), for all $t\geqslant 0$ and $%
0\leqslant s<t,$ then the corresponding optimal wealth process $X^{\ast
}=\left( X_{t}^{\ast }\right) _{t\geqslant 0}$ is given by 
\begin{align}
X_{t}^{\ast }=\left( u_{t}^{\prime }\right) ^{-1}\left( \lambda _{s,t}^{\ast
}\left( X_{s}^{\ast }\right) \mathbb{E}\left[ \rho _{s,t}^{1-\gamma }\right]
\rho _{s,t}^{\gamma }\right) =\left( u_{t}^{\prime }\right) ^{-1}\left(
u_{0}^{\prime }(x)\mathbb{E}\left[ \rho _{t}^{1-\gamma }\right] \rho
_{t}^{\gamma }\right) ,\quad 0\leqslant s<t.  \label{optimal-wealth-general}
\end{align}%
Furthermore, the Lagrangian multiplier $\lambda^{*}_{s,t}$ corresponding to the budget constraint $\mathbb{E}\left[ \rho_{s,t} X^*_t \vert \mathcal{F}_{s} \right] = X^*_s$, where $X^*_t$ is given as in (\ref{optimal-wealth-general}), satisfies $\lambda^{*}_{s,t} (X^*_s) = u_s^{\prime} (X^*_s)$ and the optimal investment policy is given by
\begin{align}\label{eq:OptimalStrategyForwardRDU}
\pi^*_t & = - \gamma \sigma_t^{-1} \lambda_t \frac{u_t^{\prime}\left( X^*_t \right)}{u_t^{\prime \prime } \left( X_t^* \right)}   .
\end{align}
ii) if $w_{s,t},$ $t\geqslant 0$ and $0\leqslant s<t,$ satisfies (\ref%
{distortion-forward-degenerate}), then $X_{t}^{\ast }=x,$ $t\geqslant 0$ and the optimal policy is $\pi^{\ast}_t = 0$, $t \geqslant 0$.
\end{proposition}

The above results give several valuable insights on the nature of the
candidate probability distortion processes. Firstly, we see that forward probability distortions satisfy a ``bifurcation" result, in that it is either
the case that equality (\ref{distortion-forward}) holds for all times and
all $p\in \left[ 0,1\right] ,$ or inequality (\ref%
{distortion-forward-degenerate}) holds throughout. The latter is a
degenerate case, as it induces zero optimal investment in all risky assets. 

The non-degenerate case, given by (\ref{distortion-forward}), has striking
similarities with a popular distortion function used in the insurance literature.
Specifically, it resembles the distortion $w(p)=\Phi \left( \Phi ^{-1}\left(
p\right) +\alpha \right) ,$ for some $a\in \mathbb{R},$ which was proposed
by Wang (see \cite{Wang00}). However, in (\ref{distortion-forward}) the
analogous ``displacement" term is neither exogenous to the market (as the coefficient $a$ is) nor static. Rather, it depends on both the investor's probability distortion parameter $\gamma $ and the market performance, as measured by the term $\sqrt{\int_{s}^{t}\Vert \lambda _{r}\Vert ^{2}dr}$ with $\lambda $ being the market price of risk process. This is intuitively pleasing as forward performance criteria are expected to follow the market
changes in ``real-time", and we see that $w_{s,t}$ does exactly this.  

We also see that the dynamic displacement $\left( \gamma -1\right) \sqrt{%
\int_{s}^{t}\Vert \lambda _{r}\Vert ^{2}dr}$ is positive (negative) if $%
\gamma >1$ ($\gamma <1),$ while the case $\gamma =0$ corresponds to no
probability distortion.

\bigskip

To give an economic interpretation of the distortion parameter $\gamma$ we recall the notion of \textit{pessimism} introduced in \cite{Quiggin93}. For a RDU representation $V$ as given in (\ref{eq:RDU-PreferenceValue}), with utility function $u$, distortion function $w$ and a prospect $X$, one can define the \textit{certainty equivalent} $CE(X)$ of $X$,  by $CE(X) := u^{-1} \left( V(X) \right)$ and \textit{risk-premium} $\Delta (X)$ of $X$ by $\Delta (X) := \mathbb{E} [X] - CE(X)$ exactly as in expected utility theory. Under RDU however, the risk premium of $X$ can be decomposed into the \textit{pessimism premium} $\Delta_{w} (X)$ of $X$, defined by 
\begin{align*}
 \Delta_w (X) := \mathbb{E} [X] -   \int_0^\infty \xi d\left( -w (1-F_X(\xi) \right) ,
 \end{align*} 
and the \textit{outcome premium} $\Delta_{u,w} (X)$ of $X$, defined by
\begin{align*}
\Delta_{u,w} (X) : = \int_0^\infty \xi d\left( -w (1-F_X(\xi) \right)  - CE(X).
\end{align*}
Indeed, one clearly has that $\Delta (X) = \Delta_w (X) + \Delta_{u,w} (X)$. We refer to \cite{GhossoubHeXX} for an extended discussion and further results on comparative RDU preferences. 
\begin{definition}
Let $V,V_1,V_2$ be RDU representations as in (\ref{eq:RDU-PreferenceValue}) with utility functions $u,u_1,u_2$ and distortion functions $w,w_1,w_2$, respectively. Then, 
\begin{enumerate}
\item[i)] $V$ is said to be pessimistic if for any $X$ with bounded support, $\Delta_w (X) \geq 0$. 
\item[ii)] $V_1$ is said to be more pessimistic than $V_2$ if for any $X$ with bounded support, $\Delta_{w_1} (X) \geqslant \Delta_{w_2} (X)$.
\end{enumerate}
\end{definition}

The following proposition shows how the distortion parameter $\gamma$ reflects the investor's attitude as objective $\left( \gamma
=1\right),$ pessimistic $\left( \gamma <1\right) $ or optimistic $\left( \gamma >1\right).$

\begin{proposition}\label{prop:Pessimism}
Let $V,V_1,V_2$ be RDU representations as in (\ref{eq:RDU-PreferenceValue}) with utility functions $u,u_1,u_2$ and distortion functions $w,w_1,w_2$ given by (\ref{distortion-forward}) with distortion parameters $\gamma, \gamma_1, \gamma_2$ respectively. Then the following holds:
\begin{enumerate}
\item[i)] $V$ is pessimistic if and only if $\gamma \leqslant 1$.
\item[ii)] $V_1$ is more pessimistic than $V_2$ if and only if $\gamma_1 \leqslant \gamma_2$.
\end{enumerate}
\end{proposition}

While most commonly used probability distortion functions, such as the ones introduced in \cite{TverskyKahneman92}, \cite{TverskyFox95} or \cite{Prelec98}, do not satisfy the Jin-Zhou monotonicity condition when paired with a lognormal pricing kernel, the proof of Theorem \ref{thm:WeightingNecessary} shows that an endogenously determined probability distortion function of a forward rank-dependent performance criterion automatically satisfies this condition. In other words, while general classical (backward) RDU optimization problems are typically hard to solve and rely on concavification techniques due to joint nonlinearities of the risk preferences and non-linear averaging, the endogenous determination of the probability distortion by means of forward criteria provides additional structure in terms of the Jin-Zhou monotonicity. This, in turn, leads to a simpler expression for the optimal wealth process as described in Proposition \ref{prop:OptimalWealthLagrangianMultiplier}.

Interestingly, \cite{XiaZhou16} also find that the Jin-Zhou monotonicity condition is automatically satisfied for a representative agent when it is the pricing kernel which is endogenously determined through an equilibrium condition of an Arrow-Debreu economy.

Finally, we comment on the form of the optimal wealth process as it plays a
pivotal role in developing the upcoming construction approach. In the
degenerate case, we easily deduce that $X_{t}^{\ast }=x,$ $t\geqslant 0.$ Note
that the no participation effect occurs even if $\lambda _{t}>0,$ $t\geqslant 0,$
as assumed herein. 

For the non-degenerate case, we see that $X_{t}^{\ast }$ takes a form that
resembles the one in the classical setting but under a \textit{different}
measure, as manifested by the term $\mathbb{E}\left[ \rho _{t}^{1-\gamma }%
\right] \rho _{t}^{\gamma }$ in the second equality in (\ref%
{optimal-wealth-general}). This motivated us to introduce a new measure,
which in turn guided us to develop a connection with the existing
deterministic, time-monotone forward criteria in an auxiliary market. We present these
results in the next section. 

\section{Construction of forward rank-dependent criteria}

\label{sec:ForwardRDUDistortedMeasure}

This section contains the main result herein. It provides a direct
connection between forward rank-dependent performance criteria $\left(
\left( u_{t}\right) _{t\geqslant 0},\left( w_{s,t}\right) _{0\leqslant s<t}\right) $
in the original market and deterministic, time-monotone forward criteria $\left(
U_{t}\right) _{t\geqslant 0}$ in an auxiliary market. 

\subsection{The auxiliary market }\label{subsec:AuxiliaryMarket}

For a fixed number $\gamma \geqslant 0$, we let $\mathbb{P}_{\gamma }$ be the
unique probability measure on $(\Omega ,\mathcal{F})$ satisfying, for each $%
t\geqslant 0,$ 
\begin{align}
\left. \frac{d\mathbb{P}_{\gamma }}{d\mathbb{P}}\right \vert _{\mathcal{F}%
_{t}}=\frac{\rho _{t}^{1-\gamma }}{\mathbb{E}[\rho _{t}^{1-\gamma }]}.
\label{gamma-measure}
\end{align}%
Such a probability measure exists, and is unique and equivalent to $\mathbb{P%
}$ on $\left( \Omega ,\mathcal{F}_{t}\right) ,$ for each $t\geqslant 0$, see,
e.g., \cite{KaratzasShreve98}. We call $\mathbb{P}_{\gamma }$ the $\gamma $%
\textit{-distorted probability measure}, see also \cite{MaWongZhangXX}. 

In turn, the price processes (cf. (\ref{stock})) solve 
\begin{align}
dS_{t}=\sigma _{t}S_{t}\left( \lambda _{\gamma ,t}dt+dW_{\gamma ,t}\right)
,\quad t\geqslant 0,  \label{stock-gamma}
\end{align}%
where 
\begin{align}
\lambda _{\gamma ,t}:=\gamma \lambda _{t}  \label{gamma-MPR}
\end{align}%
and 
\begin{align}
W_{\gamma ,t}:=\left( 1-\gamma \right) \int_{0}^{t}\lambda _{s}ds+W_{t}
\label{W-gamma}
\end{align}%
is a Brownian motion under $\mathbb{P}_{\gamma }$.

We now consider an auxiliary market consisting of the riskless bond (with
zero interest rate) and $N$ stocks whose prices evolve as in (\ref%
{stock-gamma}) above. We will refer to this as the $\gamma $\textit{-distorted market. 
}It is complete and its pricing kernel, denoted by $\rho _{\gamma ,t},$ is
given by 
\begin{align}
\rho _{\gamma ,t}=\rho _{t}\frac{d\mathbb{P}}{d\mathbb{P}_{\gamma }}=\rho
_{t}^{\gamma }\mathbb{E}[\rho _{t}^{1-\gamma }].  \label{rho-gamma}
\end{align}%
In this auxiliary market, we recall the associated time-monotone forward performance criteria, denoted by $U_{t}\left( x\right).$ It is given by 
\begin{align}
U_{t}\left( x\right) =v(x,A_{\gamma ,t})\text{ \  \  \ with \  \ }A_{\gamma
,t}:=\int_{0}^{t}\Vert \lambda _{\gamma ,s}\Vert ^{2}ds.  \label{v-gamma}
\end{align}%
The function $v(x,t)$ solves, for $x\geqslant 0,$ $t\geqslant 0,$%
\begin{align*}
v_{t}=\frac{1}{2}\frac{v_{x}^{2}}{v_{xx}},
\end{align*}%
and $v\left( x,0\right) $ must be of the form $\left( v^{\prime }\right)
^{-1}\left( x,0\right) =\int_{0^{+}}^{\infty }x^{-y}\mu \left( dy\right) ,$
where $\mu $ is a positive finite Borel measure. It also holds that

\begin{align}
v\left( x,t\right) =-\frac{1}{2}\int_{0}^{t}e^{-h^{-1}(x,s)+\frac{s}{2}%
}ds+\int_{0}^{x}e^{-h^{-1}\left( z,0\right) }dz,  \label{v-h}
\end{align}%
with $h(z,t),$ $z\in \mathbb{R},$ $t\geqslant 0,$ given by 
\begin{align}
h\left( z,t\right) :=\int_{0^{+}}^{\infty }e^{zy-\frac{1}{2}y^{2}t}\mu (dy).
\label{h-function}
\end{align}%
We refer the reader to \cite{MusielaZariphopoulou10a} for an extensive
exposition of these results as well as detailed assumptions on the
 underlying measure $\mu .$ 

\bigskip 

We are now ready to present the main result herein, which connects the
forward rank-dependent criteria in the original market with the
deterministic, time-monotone forward criteria in the $\gamma$-distorted market and provides a \textit{construction} method for forward rank-dependent performance criteria.

\begin{theorem}\label{thm:ClassicalForwardUtilityEquivalentMeasure}
Let $\gamma \geq 0$. If $\left( u_t \right)_{t \geq 0}$ is a deterministic, time-monotone forward performance criterion in the $\gamma$-distorted market and the family of probability distortions $\left( w_{s,t} \right)_{0 \leq s < t}$ is defined by (\ref{distortion-forward}), then $\left( \left( u_{t} \right)_{t \geq 0} , \left(w_{s,t} \right)_{0 \leq s < t} \right)$ is a forward rank-dependent performance criterion.

Conversely, let $\left( \left( u_{t} \right)_{t \geq 0} , \left(w_{s,t} \right)_{0 \leq s < t} \right)$ be a forward rank-dependent performance criterion. Then there exists a $\gamma \geq 0$ such $\left( u_t \right)_{t\geq 0}$ is a deterministic, time-monotone forward performance criterion in the $\gamma$-distorted market.

Forward rank-dependent performance criteria can thus be constructed as follows. Let $\gamma \geqslant 0$ and consider an initial datum of
the form $\left( u^{\prime }\right) ^{-1}\left( x,0\right)
=\int_{0^{+}}^{\infty }x^{-y}\mu \left( dy\right) $. Let $h(z,t)$ and $%
v\left( x,t\right) $ be given by (\ref{h-function})\ and (\ref{v-h}). Then, the pair $\left( \left( u_{t}\right) _{t\geqslant 0},\left(
w_{s,t}\right) _{0\leqslant s<t}\right) ,$ with  
\begin{align}
u_{t}(x):=v\left( x,\gamma ^{2}\int_{0}^{t}\Vert \lambda _{s}\Vert
^{2}ds\right) \  \text{\  \  \ and }\ w_{s,t}(p):=\frac{1}{\mathbb{E}\left[
\rho _{s,t}^{1-\gamma }\right] }\int_{0}^{p}\left( \left( F_{s,t}^{\rho
}\right) ^{-1}(q)\right) ^{1-\gamma }dq  \label{u-w-constructed}
\end{align}%
is a forward rank-dependent performance criterion.
\end{theorem}

\bigskip

We stress that, while the necessary conditions on the probability distortion
function in Theorem \ref{thm:WeightingNecessary} have been established
independently of the utility function process $\left( u_{t}\right) _{t\geqslant 0}
$, both processes ($u_{t},w_{s,t}$) depend on the distortion parameter $%
\gamma $ as (\ref{u-w-constructed}) indicates. Indeed, $\gamma >0$ manifests
itself \textit{both} as a parameter in the probability distortion function
and in the rescaled time argument for the utility function,through the
process $A_{\gamma ,t}$.

As $\gamma \downarrow 0,$ then $\lim_{\gamma \downarrow 0}A_{\gamma ,t}=0,$
for all $t\geqslant 0,$ and in turn $u_{t}(x)=u_{0}(x)=v\left( x,0\right) .$ This is expected, as when $\gamma =0,$ the risky asset prices in the $\gamma-$distorted market become martingales (cf. (\ref{stock-gamma})\ and thus no participation is expected. Indeed, the $\gamma$-distorted measure $\mathbb{P}_{\gamma}$ coincides with the risk-neutral measure when $\gamma = 0$.

\begin{proposition}\label{prop:ConstructedOptimalWealthStrategy}
Let $\gamma >0$ be the
investor's distortion parameter and $h\left( z,t\right) $ as in (\ref{h-function}). Then, the associated optimal wealth, $\left(
X_{t}^{\ast }\right) _{t\geqslant 0}$ and investment policy $\left(
\pi _{t}^{\ast }\right) _{t\geqslant 0}$ corresponding to the forward rank-dependent performance criterion as constructed in (\ref{u-w-constructed}) are given, respectively, by
\begin{align}
X_{t}^{\ast }=h\left( h^{-1}\left( x,0\right) +\gamma \int_{0}^{t}\Vert
\lambda _{s}\Vert ^{2}ds+\gamma \int_{0}^{t}\lambda _{s}dW_{s},\gamma
^{2}\int_{0}^{t}\Vert \lambda _{s}\Vert ^{2}ds\right)   \label{wealth-gamma}
\end{align}
and
\begin{align}
\pi _{t}^{\ast }=\gamma \sigma _{t}^{-1}\lambda _{t}h_{x}\left( h^{-1}\left( x,0\right) +\gamma \int_{0}^{t}\Vert \lambda _{s}\Vert ^{2}ds+\gamma
\int_{0}^{t}\lambda _{s}dW_{s},\gamma ^{2}\int_{0}^{t}\Vert \lambda
_{s}\Vert ^{2}ds\right) .  \label{pi-gamma}
\end{align}
ii) Let $\gamma =0.$ Then, $X_{t}^{\ast }=x$ and $\pi
_{t}^{\ast }=0,$  for all $t\geqslant 0.$
\end{proposition}

We remind the reader that (\ref{wealth-gamma}) and (\ref{pi-gamma}) offer an alternative
expression for the optimal wealth and policy, already derived with different arguments
in (\ref{optimal-wealth-general}) and (\ref{eq:OptimalStrategyForwardRDU}).

\bigskip

\textit{\bigskip }

Next, we provide examples where the underlying measure $\mu $ is a single Dirac or sum of two Dirac functions.

\begin{example}
\label{ex:CRRA} i) Let $u_{0}(x)=\frac{1}{1-\alpha }x^{1-\alpha },$ $\alpha
\neq 1.$

Equivalently, $\mu (dy)=\delta _{1/\alpha }$ and, in turn, $h\left(
x,t\right) =e^{\frac{x}{\alpha }-\frac{1}{2\alpha ^{2}}t}$ and $v(x,t)=\frac{%
1}{1-\alpha }x^{1-\alpha }e^{\frac{1}{2}(1-\frac{1}{\alpha })t}.$ Let $%
\gamma >0.$ Then, the pair $\left( \left( u_{t}\right) _{t\geqslant 0},\left(
w_{s,t}\right) _{0\leqslant s<t}\right) $ defined, for $t\geqslant 0$ and $0\leqslant s<t,$
as 
\begin{align*}
u_{t}\left( x\right) =\frac{1}{1-\alpha }x^{1-\alpha }e^{\frac{1}{2}(1-\frac{%
1}{\alpha })\gamma ^{2}\int_{0}^{t}\Vert \lambda _{s}\Vert ^{2}ds}\text{ \  \
\ and \  \  \ }w_{s,t}(p):=\frac{1}{\mathbb{E}\left[ \rho _{s,t}^{1-\gamma }%
\right] }\int_{0}^{p}\left( \left( F_{s,t}^{\rho }\right) ^{-1}(q)\right)
^{1-\gamma }dq
\end{align*}%
is a forward rank-dependent criterion. Furthermore, from (\ref{wealth-gamma}
) and (\ref{pi-gamma}) we deduce that
\begin{align}
X_{t}^{\ast }=xe^{\frac{\gamma }{\alpha }\left( 1-\frac{\gamma }{2\alpha }%
\right) \int_{0}^{t}\Vert \lambda _{s}\Vert ^{2}ds+\frac{\gamma }{\alpha }%
\int_{0}^{t}\lambda _{s}dW_{s}}\text{ \  \  \ and \  \  \  \  \ }\pi _{t}^{\ast }=%
\frac{\gamma }{\alpha }\sigma _{t}^{-1}\lambda _{t}X_{t}^{\ast }.
\label{wealth-power-1}
\end{align}%
Direct calculations also yield that, for $t\geqslant 0,$ 
\begin{align}
X_{t}^{\ast }=x\frac{\rho _{t}^{-\frac{\gamma }{\alpha }}}{\mathbb{E}\left[
\rho _{t}^{1-\frac{\gamma }{\alpha }}\right] },  \label{wealth-power}
\end{align}%
with the pricing kernel given in (\ref{kernel-single}).

ii) Let $u_{0}\left( x\right) =\log x.$ Then $\mu (dy)=\delta _{1}$ and, in
turn, $h\left( x,t\right) =e^{x-\frac{1}{2}t}$ and $v(x,t)=\log x-\frac{1}{2}%
t.$

For $\gamma >0,$ the pair $\left( \left( u_{t}\right) _{t\geqslant 0},\left(
w_{s,t}\right) _{0\leqslant s<t}\right) $ defined, for $t\geqslant 0$ and $0\leqslant s<t,$
as 
\begin{align*}
u_{t}\left( x\right) =\log x-\frac{1}{2}\gamma ^{2}\int_{0}^{t}\Vert \lambda
_{s}\Vert ^{2}ds\text{\  \  \  \  \ and \  \  \ }w_{s,t}(p):=\frac{1}{\mathbb{E}%
\left[ \rho _{s,t}^{1-\gamma }\right] }\int_{0}^{p}\left( \left(
F_{s,t}^{\rho }\right) ^{-1}(q)\right) ^{1-\gamma }dq
\end{align*}%
is a forward rank-dependent criterion.

Furthermore, from (\ref{wealth-gamma}) and (\ref{pi-gamma}) 
\begin{align*}
X_{t}^{\ast }=xe^{\gamma \left( 1-\frac{\gamma }{2}\right) \int_{0}^{t}\Vert
\lambda _{s}\Vert ^{2}ds+\gamma \int_{0}^{t}\lambda _{s}dW_{s}}\text{ \  \  \
and \  \  \ }\pi _{t}^{\ast }=\gamma \sigma _{t}^{-1}\lambda _{t}X_{t}^{\ast }.
\end{align*}
\end{example}

\begin{example}\label{ex:SumTwoDiracs}
Let 
$
u_0(x) = \frac{2^{-\theta}}{\theta (1+\theta)}  \left( \sqrt{4 x  +1} - 1 \right)^{\theta} \left( \theta \sqrt{4 x  + 1} +1 \right),
$
 $0 < \theta < 1$.

Equivalently, the underlying measure is $\mu (dy)=\delta _{\frac{1}{1-\theta}} + \delta _{\frac{2}{1-\theta}}$ and therefore 
\begin{align*}
h(x,t) = e^{\frac{1}{1-\theta }x-\frac{1}{2}\frac{1}{\left( 1-\theta \right)
^{2}}t}+e^{\frac{2}{1-\theta }x-\frac{2}{\left( 1-\theta \right)
^{2}}t}
\end{align*}
 and 
\begin{align*}
v(x,t) = \frac{2^{-\theta}}{\theta (1+\theta)} e^{\frac{t}{2}\left( 1 - \frac{3}{1-\theta} + \frac{2}{(1-\theta)^2} \right)} \left( \sqrt{4 x e^{-\frac{t}{(1-\theta)^2}} +1} - 1 \right)^{\theta} \left( \theta \sqrt{4 x e^{- \frac{t}{(1-\theta)^2}} + 1} +1 \right)
 \end{align*} 
Let $\gamma > 0$.  Then, the pair $\left( \left( u_{t}\right) _{t\geqslant 0},\left(
w_{s,t}\right) _{0\leqslant s<t}\right) $ defined, for $t\geqslant 0$ and $0\leqslant s<t,$
by
\begin{align*}
u_{t}\left( x\right) & = \frac{2^{-\theta}}{\theta (1+\theta)} e^{\frac{ \gamma ^{2}}{2}\left( 1 - \frac{3}{1-\theta} + \frac{2}{(1-\theta)^2} \right) \int_{0}^{t}\Vert \lambda
_{s}\Vert ^{2}ds } \\
& \qquad \times \left( \sqrt{4 x e^{-\frac{ \gamma ^{2}}{(1-\theta)^2} \int_{0}^{t}\Vert \lambda
_{s}\Vert ^{2}ds } +1} - 1 \right)^{\theta} \left( \theta \sqrt{4 x e^{- \frac{ \gamma ^{2}}{(1-\theta)^2}\int_{0}^{t}\Vert \lambda
_{s}\Vert ^{2}ds} + 1} +1 \right)
\end{align*}
and 
\begin{align*}
w_{s,t}(p):=\frac{1}{\mathbb{E}\left[ \rho _{s,t}^{1-\gamma }%
\right] }\int_{0}^{p}\left( \left( F_{s,t}^{\rho }\right) ^{-1}(q)\right)
^{1-\gamma }dq
\end{align*}
is a forward rank-dependent criterion. Furthermore, from (\ref{wealth-gamma}
) and (\ref{pi-gamma}) we deduce that 
\begin{align}
\begin{split}
X_{t}^{\ast } &  =  \frac{1}{2} \left( \sqrt{4 x + 1} -1 \right) e^{  \frac{\gamma}{1-\theta } \left( 1 - \frac{\gamma}{2(1-\theta)} \right) \int_{0}^{t}\Vert \lambda _{s}\Vert^{2}ds + \frac{\gamma}{1-\theta } \int_{0}^{t}\lambda _{s}dW_{s}  }\\
& \quad + \frac{1}{4} \left( \sqrt{4 x + 1} -1 \right)^2 e^{ \frac{2 \gamma}{1-\theta } \left( 1 - \frac{\gamma}{1-\theta} \right) \int_{0}^{t}\Vert \lambda _{s}\Vert ^{2}ds + \frac{2 \gamma}{1-\theta } \int_{0}^{t}\lambda _{s}dW_{s}}
\end{split}
\label{wealth-ex2}
\end{align}
and 
\begin{align}
\begin{split}
\pi _{t}^{\ast } &=  \frac{\gamma}{2(1-\theta)} \sigma _{t}^{-1}\lambda _{t} \bigg( \left( \sqrt{4 x + 1} -1 \right) e^{  \frac{\gamma}{1-\theta } \left( 1 - \frac{\gamma}{2(1-\theta)} \right) \int_{0}^{t}\Vert \lambda _{s}\Vert^{2}ds + \frac{\gamma}{1-\theta } \int_{0}^{t}\lambda _{s}dW_{s}  }\\
& \hspace*{3.7cm} +  \left( \sqrt{4 x + 1} -1 \right)^2 e^{ \frac{2 \gamma}{1-\theta } \left( 1 - \frac{\gamma}{1-\theta} \right) \int_{0}^{t}\Vert \lambda _{s}\Vert ^{2}ds + \frac{2 \gamma}{1-\theta } \int_{0}^{t}\lambda _{s}dW_{s}} \bigg).
\end{split} \label{strategy-ex2}
\end{align}
\end{example}

\section{Relations with the dynamic utility approach}

\label{sec:DynamicUtilityApproach}

Motivated by the construction of time-consistent rank-dependent criteria we here revisit the classical (backward) RDU optimization problem. Specifically, we follow the \textit{dynamic utility
approach} developed in \cite{KarnamMaZhang17} and explore whether it is possible to derive a family of dynamic RDU optimization problems under which the initial investment policy remains optimal over time. In a recent work related to this paper, \cite{MaWongZhangXX} utilize this approach to derive a time-consistent conditional expectation under probability distortion. 

We emphasize that, for this section only, we deviate from the theme of
forward criteria in that we consider a classical rank-dependent utility
maximization problem of the form 
\begin{align}  \label{prob:InitialRDU}
\max_{\pi \in \mathcal{A}} & \quad \int_0^\infty u_{0,T}(\xi) d \left(- w_{0,T}
\left( 1- F_{X^\pi_{T}}(\xi) \right) \right) 
\end{align}
with $dX_{r}^{x,\pi }=\pi _{r}^{\prime }\mu _{r}dr+\pi _{r}^{\prime }\sigma _{r}dW_{r},$ $r\in \lbrack 0,T]$ and $X_{0}^{x,\pi }=x >0$
for a fixed time horizon $T > 0$, utility function $u_{0,T} \in \mathcal{U}$ and
probability distortion function $w_{0,T} \in \mathcal{W}$. We assume that there
exists an optimal policy $\pi^*$ to problem (\ref{prob:InitialRDU}) and
make the following definition of \textit{dynamic rank-dependent utility
processes}.

\begin{definition}
A family of utility functions $u_{t,T} \in \mathcal{U}$, $t \in (0,T)$, and
probability distortion functions $w_{t,T} \in \mathcal{W}$, $t \in (0,T)$ is
called a dynamic rank-dependent utility process for $u_{0,T} \in \mathcal{U}$
and $w_{0,T} \in \mathcal{W}$ over the time horizon $T > 0$ if $\lim_{t \searrow
0} u_{t,T} = u_{0,T}$, $\lim_{t \searrow 0} w_{t,T} = w_{0,T}$ and the optimal policy $\pi^*$ for (\ref{prob:InitialRDU}) also
solves,  for any $t \in (0,T)$, 
\begin{align}  \label{prob:DynamicRDU}
\max_{\pi \in \mathcal{A} (\pi^*,t)} & \quad \int_0^\infty u_{t,T}(\xi) d
\left(- w_{t,T} \left( 1- F_{X^{x,\pi}_{T} \vert \mathcal{F}_t} (\xi)\right)
\right) 
\end{align}
with $dX_{r}^{x,\pi }=\pi _{r}^{\prime }\mu _{r}dr+\pi _{r}^{\prime }\sigma _{r}dW_{r},$ $r\in \lbrack 0,T]$ and $X_{0}^{x,\pi }=x >0$.
\end{definition}

Our definition of dynamic rank-dependent utility processes relies on the existence of an optimal policy for the initial problem (\ref{prob:InitialRDU}). \cite{KarnamMaZhang17} do not rely on this assumption and being able to determine the value of a time-inconsistent stochastic control problem without the assumption of an optimal control is indeed one of their main contributions. However, for our specific problem of rank-dependent utility maximization in a complete, continuous-time financial market with deterministic coefficients, conditions for the existence of an optimal policy are well known and not restrictive, cf. \cite{XiaZhou16}, \cite{Xu16} or \cite{HeKouwenbergZhou17}.

We also deviate from \cite{KarnamMaZhang17} in that we restrict the
objective of the dynamic family of problems (\ref{prob:DynamicRDU}) to
belong to the same class of preferences as the initial problem (\ref%
{prob:InitialRDU}), namely rank-dependent utility preferences. In \cite{KarnamMaZhang17} on the other hand, the objective is allowed to vary more freely. However, we believe that within the dynamic utility approach, it is an interesting mathematical and economic question whether one is able to construct a dynamic utility belonging to the same class of preference functionals as the initial preferences and so this is exactly the question we want to address. There have been some positive results in this regard for generally time-inconsistent mean-risk portfolio optimization problems; see \cite{CuiLiWangZhu12} for the mean-variance and \cite{StrubLiCuiGaoXX} for the mean-CVaR problem.

\bigskip

\cite{MaWongZhangXX} introduce the notion of a \textit{dynamic distortion function} under which the distorted, nonlinear conditional expectation is time-consistent, in the sense that the tower-property holds. Specifically, they consider an It\^{o} process described by the stochastic differential equation 
\begin{align}  \label{eq:ItoProcess}
Y_{t} = y_{0} + \int_{0}^{t} b(s,Y_s) ds + \int_0^{t} \sigma (s, Y_{s}) d
W_{s}, \quad 0 \leqslant t \leqslant T
\end{align}
and a given family of probability distortion functions $\left( w_{0,t}
\right)_{0 \leqslant t \leqslant T}$, where $w_{0,t}$ applies over $[0,t]$. Under
technical conditions, they are able to derive a family of (random)
probability distortion functions $\left( w_{s,t} \right)_{0 \leqslant s \leqslant t \leqslant T}$ such that $w_{s,t}$ is $\mathcal{F}_{s}$-measurable, for any $0 \leqslant s \leqslant t \leqslant T$, and the tower-property 
\begin{align*}
\mathcal{E}_{r,t} [g(Y_{t})] = \mathcal{E}_{r,s} \left[ \mathcal{E}_{s,t} %
\left[ g (Y_t) \right] \right], \quad 0 \leqslant r \leqslant s \leqslant t \leqslant T,
\end{align*}
holds for any continuous, bounded, increasing and nonnegative function $g$, where $%
\mathcal{E}_{s,t}$ denotes the nonlinear conditional expectation 
\begin{align*}
\mathcal{E}_{s,t} [\xi] = \int_0^\infty w_{s,t} \left( \mathbb{P} \left[ \xi
\geqslant x \right] \right) dx.
\end{align*}
We remark that the tower-property plays an important role in the theory on dynamic risk measures, see, e.g., \cite{BieleckiCialencoPitera17} for a survey, and refer to \cite{MaWongZhangXX} for further applications and discussions.

The important difference between the portfolio optimization problem we study here and the setting of \cite{MaWongZhangXX} is that the It\^{o} process described through the stochastic differential equation (\ref{eq:ItoProcess}) is given and fixed. In particular, there is no control or investment policy. Moreover, the construction of the family of probability distortion functions $\left( w_{s,t} \right)_{0 \leqslant s \leqslant t \leqslant T}$ in \cite%
{MaWongZhangXX} depends on the drift $b$ and volatility $\sigma$ in (\ref{eq:ItoProcess}), cf. Theorem 5.2 therein. In our setting, on the other hand, the drift and volatility of the wealth process are controlled by the investment policy $\pi$.\newline

It follows as a corollary to our results on forward rank-dependent
performance processes that, when the utility function $u_{t,T}=u_{0,T}$ for any $t\in (0,T)$ and the optimal policy invests is not degenerate,
we can construct a dynamic rank-dependent utility process if and only if the
probability distortion function belongs to the class of \cite{Wang00}. The
following theorem shows that this result remains valid even if one allows
both the utility function $u_{t,T}$ and probability distortion function $w_{t,T}$ to depend on the initial time of the investment.

\begin{theorem}
\label{thm:DynamicRDU} Consider a fixed time-horizon $T$, utility function $u_{0,T}\in \mathcal{U}$ and probability distortion function $w_{0,T}\in \mathcal{W}$, and suppose that the optimal solution to (\ref{prob:InitialRDU}) exists. Depending on the probability distortion function $w_{0,T}$, we have the following two cases:

i) If 
\begin{align} \label{eq:DynamicDistortionDegenerate}
w_{0,T}(p) \geqslant \mathbb{E}\left[ \rho _{0,T}\mathbf{1}_{\left \{ \rho _{0,T}\text{ }%
\leqslant \left( F_{0,T}^{\rho }\right) ^{-1}(p)\right \} }\right] ,\quad p\in
\lbrack 0,1],
\end{align}
then a family of utility functions $u_{t,T}\in \mathcal{U}$, $t\in (0,T)$, together with a family of probability distortion functions $w_{t,T}\in 
\mathcal{W}$, $t\in ( 0,T )$, is a dynamic rank-dependent utility
process for $u_{0,T}$ and $w_{0,T}$ if and only if the family of probability distortion functions satisfies 
\begin{align*}
w_{t,T}(p)\geqslant \mathbb{E}\left[ \rho _{t,T}\mathbf{1}_{\left \{ \rho _{t,T}%
\text{ }\leqslant \left( F_{t,T}^{\rho }\right) ^{-1}(p)\right \} }\right] ,\quad
p\in \lbrack 0,1],
\end{align*}%
for any $t\in \lbrack 0,T)$.

ii) If (\ref{eq:DynamicDistortionDegenerate}) does not hold, then a family of utility
functions $u_{t,T} \in \mathcal{U}$, $t \in (0,T)$ and probability
distortion functions $w_{t,T} \in \mathcal{W}$, $t \in (0,T)$ with $\lim_{t
\searrow 0} u_{t,T}  = u_{0,T}$ and $\lim_{t \searrow 0} w_{t,T} =
w_{0,T} $ is a dynamic rank-dependent utility process for $u_{0,T}$ and $%
w_{0,T} $ if and only if there is a deterministic process $\gamma_t \geqslant 0$, $t
\in [0,T)$, continuous at zero and such that 
\begin{align}\label{eq:DynamicWeighting}
w_{t,T} (p) = \frac{1}{\mathbb{E}\left[ \rho_{t,T}^{1-\gamma_t}\right]}
\int_0^p \left( \left( F^\rho_{t,T} \right)^{-1} (q) \right)^{1-\gamma_t} dq = \Phi \left( \Phi ^{-1}\left( p\right) +\left( \gamma_t -1\right) 
\sqrt{\int_{t}^{T}\Vert \lambda _{r}\Vert ^{2}dr}\right)
\end{align}
for any $t \in [0,T)$, and the measure of risk-aversion of the dynamic
utility function satisfies 
\begin{align}  \label{eq:RiskAversionDynamicRDU}
- \frac{u_{t,T}^{\prime \prime }(x) }{u_{t,T}^{\prime }(x)} = - \frac{\gamma_t}{\gamma_0} \frac{u_{0,T}^{\prime \prime }(x) }{u_{0,T}^{\prime }(x)}
\end{align}
for any $t \in [0,T)$ and $x > 0$.
\end{theorem}

Theorem \ref{thm:DynamicRDU} shows in particular that, when there is
some non-zero investment in the risky asset, an extension of the construction of
dynamic distortion functions of \cite{MaWongZhangXX} to controlled processes
is in general only possible if the initial probability distortion function
belongs to the family introduced in \cite{Wang00}.

In order to maintain time-consistency, the utility function and probability distortion function must be coordinated with each other at different times through the relationship (\ref{eq:RiskAversionDynamicRDU}). This dynamic constraint connects
the risk-aversion of the dynamic utility function with the dynamic parameter of the probability distortion function. Recall from Proposition \ref{prop:Pessimism} that the distortion parameter $\gamma_t$ reflects the investor's attitude as objective ($\gamma_t = 1$), pessimistic ($\gamma_t < 1$) or optimistic ($\gamma_t > 1$). The dynamic constraint (\ref{eq:RiskAversionDynamicRDU}) can thus be interpreted as follows: In order to be time-consistent, the investor must become more risk-averse if she becomes less pessimistic and less risk-averse if she becomes more pessimistic. Moreover, the relationship between risk-aversion and pessimism as reflected in the parameter $\gamma_t$ is linear. 

Note that, if the risk-aversion is time-invariant, then $\gamma_t = \gamma_0$, $t \in (0,T)$, implying that the effect of probability distortion (measured by the time parameter $t$) thus must decay over
time at the order of $\sqrt{T-t}$, as it follows from (\ref{eq:DynamicWeighting}). In particular, the probability distortion
effect should disappear when the remaining time approaches zero. On the
other hand, if the probability distortion function is time-invariant, we must have $\gamma_t = 1 + (\gamma_0 - 1) \sqrt{ \frac{\int_0^T \Vert \lambda_s \Vert^2 ds}{\int_t^T \Vert \lambda_s \Vert^2 ds} }$, $t \in (0,T)$. Therefore, the measure of absolute risk-aversion of the utility function must increase at
the order of $1/\sqrt{T-t}$ as $t$ approaches $T$.

\section{Conclusions}

\label{sec:Conclusions} We introduced the concept of forward rank-dependent
performance processes and thereby extended the study of forward performance
process to settings involving probability distortions. Forward
rank-dependent performance criteria are herein taken to be deterministic.
This made the problem tractable but also guided us in building a
fundamental connection with deterministic, time-monotone forward criteria in a related
market. 

We provided two alternative definitions, in terms of time-consistency and performance value preservation, respectively. We, then, provided a complete characterization of the viable probability distortion functions. Specifically, we showed that for the non-degenerate case (non-zero risky allocation) the probability distortion function resembles the one introduced by \cite{Wang00} but modified appropriately to capture the market evolution.
We also showed that it satisfies the Jin-Zhou monotonicity condition.

We further derived the optimal wealth process, which then motivated the introduction of the distorted probability measure. We in turn established the key result, namely, a one-to-one correspondence between forward rank-dependent performance processes and deterministic, time-monotone forward performance criteria in the auxiliary market (under the distorted measure). This results then allows to
build on earlier findings of \cite{MusielaZariphopoulou09,MusielaZariphopoulou10a} to characterize
forward rank-dependent performance criteria and their optimal processes.

Finally, we related our results with the dynamic utility approach of \cite{KarnamMaZhang17} and, specifically, the dynamic distortion function of \cite{MaWongZhangXX}. While \cite{MaWongZhangXX} are able to construct a dynamic distortion function which is time-consistent in the sense that the tower-property holds for a general class of initial probability distortion functions and given and fixed state process, our results show that, when the wealth processes is controlled by
the investment policy and there is investment in the risky asset, then
time-consistent investment under probability distortion is possible if and
only if the probability distortion belongs to the class of \cite{Wang00}.

Extending the deterministic case to the stochastic one is by no means
trivial, as there are both conceptual and technical challenges. Indeed, if
one would allow $u_{t}$ and/or $w_{s,t}$ to be $\mathcal{F}_{t}$-measurable,
in direct analogy to the forward performance case, then the value of a
prospect as specified in (\ref{eq:RDU-PreferenceValue}) would be a random
variable. Simply taking the expectation of this random value of the prospect
seems ad hoc. In particular, it seems unreasonable that an agent distorting
probabilities to evaluate a prospect would subsequently apply a mere linear
expectation to average the resulting value of the prospect.

There are a number of possible directions for future research. First, one might consider forward cumulative prospect theory performance criteria, which incorporate two further behavioral phenomena, namely reference dependence and loss aversion. There is a rich and active literature on how the reference point evolves in time, and the results herein indicate that the framework of forward preferences seems suitable to derive conditions under which a time-varying reference point does not lead to time-inconsistent investment policies.

A second possible direction is to consider discrete-time rank-dependent forward criteria. Indeed, much of the research in behavioral finance and economics assumes a discrete-time setting. Furthermore, considering discrete-time forward criteria already lead to valuable insights if there is no probability distortion.

Finally, it would be interesting to extend the framework of forward rank-dependent performance criteria beyond problems of portfolio selection. Possible problems could for example come from the areas of pricing and hedging, insurance, optimal contracting, real world options or in situations where there is competition between different agents.

\section*{Acknowledgments}
We are very grateful to Xunyu Zhou for suggesting the topic and his valuable comments. 

\appendix

\section{Appendix. Proofs}

\label{app:Proofs}

\subsection{Proof of Proposition \protect \ref{prop:DefinitionForwardEquivalent}} 

If $w(p)=p,p\in \lbrack 0,1]$ and $u\in \mathcal{U},$ then for any prospect $%
X$, 
\begin{align*}
\int_{0}^{\infty }u(\xi )d\left( -w\left( 1-F_{X|\mathcal{F}_{t}}(\xi
)\right) \right) =\mathbb{E}\left[ u(X)|\mathcal{F}_{t}\right] .
\end{align*}%
\qed

\subsection{Proof of Proposition \protect \ref{prop:TimeConsistency}}

Suppose that $\left( \left( u_{t}\right) _{t\geqslant 0},\left( w_{s,t}\right)
_{0\leqslant s<t}\right) $ is a  forward rank-dependent performance criterion. Assume that $\pi ^{\ast }$ is not optimal for problem (\ref{prob:TC-general})
for some $0\leqslant s<t$ and $x>0$. Let $\mathcal{A}$ and $\mathcal{A}(\pi
^{\ast },s)$ as in (\ref{admissible-set}) and (\ref{admissible-set-truncated}%
). Then, there would be a policy, say $\pi \in \mathcal{A}(\pi ^{\ast },s),
$ such that%
\begin{align*}
\int_{0}^{\infty }u_{t}(\xi )d\left( -w_{s,t}\left( 1-F_{X_{t}^{x,\pi }|%
\mathcal{F}_{s}}(\xi )\right) \right) >\int_{0}^{\infty }u_{t}(\xi )d\left(
-w_{s,t}\left( 1-F_{X_{t}^{x,\pi ^{\ast }}|\mathcal{F}_{s}}(\xi )\right)
\right) 
\end{align*}%
\begin{align*}
=u_{s}\left( X_{s}^{x,\pi ^{\ast }}\right) =u_{s}\left( X_{s}^{x,\pi
}\right) ,
\end{align*}%
on a set $A_{s}\in $ $\mathcal{F}_{s}$ with $\mathbb{P}\left[ A_{s}\right] >0
$. This however contradicts iii) of Definition \ref{def:ForwardRDU}. It then
follows that the optimal value of (\ref{prob:TC-general}) is given by $%
u_{s}\left( X_{s}^{x,\pi ^{\ast }}\right) $. \newline

Next, assume that the pair $\left( \left( u_{t}\right) _{t\geqslant 0},\left(
w_{s,t}\right) _{0\leqslant s<t}\right) $ is a time-consistent rank-dependent
performance criterion preserving the performance value. We only have to
argue that for any admissible policy $\pi $, any $t\geqslant 0$ and $0\leqslant s<t,$
and $x>0$, the inequality 
\begin{align*}
\int_{0}^{\infty }u_{t}(\xi )d\left( -w_{s,t}\left( 1-F_{X_{t}^{x,\pi }|%
\mathcal{F}_{s}}(\xi )\right) \right) \leqslant u_{s}\left( X_{s}^{x,\pi }\right) 
\end{align*}%
holds. We argue by contradiction. Suppose that there exists policy $\bar{%
\pi}\in \mathcal{A}$, time $t\geqslant 0$ and an $s$ with $0\leqslant s<t$, an $x>0
$, a set $\tilde{A}_{s}\in \mathcal{F}_{s}$ with $\mathbb{P}\left[ A_{s}%
\right] >0$ and $\varepsilon >0$ such that 
\begin{align*}
\int_{0}^{\infty }u_{t}(\xi )d\left( -w_{s,t}\left( 1-F_{X_{t}^{x,\bar{\pi}}|%
\mathcal{F}_{s}}(\xi ;\omega )\right) \right) >u_{s}\left( X_{s}^{x,\bar{\pi}%
}(\omega )\right) +\varepsilon ,\quad \omega \in \tilde{A}_{s}.
\end{align*}%
Using results on classical rank-dependent utility maximization (cf. Theorem %
\ref{thm:BackwardRDEUSolution}), the wealth process corresponding to the
policy $\pi ^{\ast }$ is given by 
\begin{align*}
X_{t}^{y,\pi ^{\ast }}=(\left( u_{t}^{\prime }\right) ^{-1}\left( \lambda
_{t}^{\ast }(y)\hat{N}_{0,t}^{\prime }\left( 1-w_{0,t}\left( F_{0,t}^{\rho
}(\rho _{0,t})\right) \right) \right) ,
\end{align*}%
for any initial wealth $y>0$, where $\hat{N}_{0,t}$ is the concave envelope
of 
\begin{align*}
N_{0,t}(z):=-\int_{0}^{w_{0,t}^{-1}(1-z)}(F_{0,t}^{\rho })^{-1}(r)dr,\quad
z\in \lbrack 0,1],
\end{align*}%
and $\lambda _{t}^{\ast }(y)$ is such that $\mathbb{E}\left[ \rho
_{t}X_{t}^{y,\pi ^{\ast }}\right] =y$.

Note that the range of $\hat{N}_{0,t}^{\prime }\left( 1-w_{0,t}\left(
F_{0,t}^{\rho }(\rho _{0,t})\right) \right) $ does not depend on the initial
wealth $y>0.$ Furthermore, using results from \cite{JinZhou08} or \cite%
{HeZhou11,HeZhou16}, we obtain that $\lambda _{0,t}^{\ast }(y)=u_{0}^{\prime
}(y)$, which has range $(0,\infty )$ due to the Inada condition.

Therefore, for any $\delta >0$, there exist $a,y_{0}>0$ such that the event 
\begin{align*}
A_{s}:=\left \{ X_{s;}^{x,\bar{\pi}}\in \lbrack a,a+\frac{\delta }{2}%
],X_{s}^{y,\pi ^{\ast }}\in \lbrack a+\frac{\delta }{2},a+\delta ]\right \}
\cap \tilde{A}_{s},
\end{align*}%
is such that $A_{s}\in \mathcal{F}_{s}$ and $\mathbb{P}\left[ A_{s}\right] >0
$. 

Next, we introduce the policy 
\begin{align*}
\vartheta _{r}(\omega ;y):=\pi _{r}^{\ast }(\omega ;y)+\left( \bar{\pi}%
_{r}(\omega ;x)-\pi _{r}^{\ast }(\omega ;y)\right) \left( \mathbf{1}%
_{[s,\infty )}(r)\times \mathbf{1}_{A_{s}}(\omega )\times \mathbf{1}%
_{y=y_{0}}\right) .
\end{align*}%
We then have that $\vartheta \in \mathcal{A}(\pi ^{\ast },s),$ since $%
X_{s}^{y_{0},\pi ^{\ast }}\geqslant X_{s}^{x,\bar{\pi}}$ on $A_{s}$.

In turn, for $\omega \in A_{s}$ we obtain%
\begin{align*}
\int_{0}^{\infty }u_{t}(\xi )d\left( -w_{s,t}\left(
1-F_{X_{t}^{y_{0},\vartheta }|\mathcal{F}_{s}}(\xi ;\omega )\right) \right)
& \geqslant \int_{0}^{\infty }u_{t}(\xi )d\left( -w_{s,t}\left( 1-F_{X_{t}^{x,\bar{%
\pi}}|\mathcal{F}_{s}}(\xi ;\omega )\right) \right) \\
& >u_{s}\left( X_{s}^{x,\bar{\pi}}(\omega )\right) +\varepsilon \\
& \geqslant
u_{s}\left( X_{s}^{y_{0},\pi ^{\ast }}(\omega )-\delta \right) +\varepsilon \\
&\geqslant u_{s}\left( X_{s}^{y_{0},\pi ^{\ast }}(\omega )\right)\\
& =\int_{0}^{\infty }u_{t}(\xi )d\left( -w_{s,t}\left( 1-F_{X_{t}^{y_{0},\pi
\ast }|\mathcal{F}_{s}}(\xi ;\omega )\right) \right) ,
\end{align*}%
where the last inequality holds for small enough $\delta $. This however
contradicts the optimality of $\pi ^{\ast }$ and we easily conclude. \qed

\subsection{Proof of Theorem \protect \ref{thm:WeightingNecessary}}

From Proposition \ref{prop:TimeConsistency} we have that $\left( \left(
u_{t}\right) _{t\geqslant 0},\left( w_{s,t}\right) _{0\leqslant s<t}\right) $ is a
time-consistent rank-dependent performance criterion preserving the
performance value. We first show that the family probability distortion
functions either satisfies the Jin-Zhou monotonicity condition, namely that
the function $\frac{(F_{s,t}^{\rho })^{-1}(\cdot )}{w_{s,t}^{\prime }(\cdot )%
}$ is nondecreasing for any $0\leqslant s<t$, or inequality (\ref%
{distortion-forward-degenerate}) holds.

Because the market is complete, $\pi ^{\ast }$ is optimal for (\ref%
{prob:TC-general}) if and only if the corresponding wealth process $X^{\ast }$
solves 
\begin{align}\label{prob:RDUIntermediateTime}
\max_{X}\int_{0}^{\infty }u_{t}(\xi )d\left( -w_{s,t}\left( 1-F_{X|\mathcal{F%
}_{s}}(\xi )\right) \right) 
\end{align}%
with $\mathbb{E}\left[ \left. \rho _{s,t}X\right \vert \mathcal{F}_{s}\right]
=X_{s}^{\ast },$ $X\geqslant 0$ and $X$ is $\mathcal{F}_{s}$-measurable,
for any $0\leqslant s<t.$

Note that (\ref{prob:RDUIntermediateTime}) is a family of random optimization problems
with different initial and terminal times, $s$ and $t,$ and state-dependent
initial state and objective function. However, this does not impose any difficulty since both the initial state and objective are known at time $s\geqslant 0$, see, e.g., Chapter 4, Section 3 in \cite{YongZhou99}.

Recall that, according to Theorem \ref{thm:BackwardRDEUSolution}, the
optimal wealth for (\ref{prob:RDUIntermediateTime}) is given by 
\begin{align*}
X_{t}^{s,t}=(u_{t}^{\prime })^{-1}\left( \lambda _{s,t}^{\ast }(X_{s}^{\ast
})\hat{N}_{s,t}^{\prime }\left( 1-w_{s,t}\left( F_{s,t}^{\rho }(\rho
_{s,t})\right) \right) \right) ,
\end{align*}%
where $\hat{N}_{s,t}$ is the concave envelope of 
\begin{align}\label{eq:FunctionN}
N_{s,t}(z):=-\int_{0}^{w_{s,t}^{-1}(1-z)}(F_{s,t}^{\rho })^{-1}(r)dr,\quad
z\in \lbrack 0,1].
\end{align}
and $\lambda _{s,t}^{\ast }(X_{s}^{\ast })>0$ is such that $\mathbb{E}\left[
\left. \rho _{s,t}X_{t}^{s,t}\right \vert \mathcal{F}_{s}\right] =X_{s}^{\ast
}$. 

The time-consistency property is satisfied if and only if $X_{t}^{s,t}\left(
X_{s}^{\ast }(x)\right) =X_{t}^{\ast }(x)$ for any $0\leqslant s<t$ and $x>0$.
This becomes  
\begin{align}\label{eq:TCFunctionalEquation}
\begin{split}
\lambda _{s,t}^{\ast }& \left( \left( u_{s}^{\prime }\right) ^{-1}\left(
\lambda _{0,s}^{\ast }(x)\hat{N}_{0,s}^{\prime }\left( 1-w_{0,s}\left(
F_{0,s}^{\rho }(\rho _{0,s})\right) \right) \right) \right) \hat{N}%
_{s,t}^{\prime }\left( 1-w_{s,t}\left( F_{s,t}^{\rho }(\rho _{s,t})\right)
\right)  \\
& =\lambda _{0,t}^{\ast }(x)\hat{N}_{0,t}^{\prime }\left( 1-w_{0,t}\left(
F_{0,t}^{\rho }(\rho _{0,t})\right) \right) .
\end{split}
\end{align}
Next, we define the auxiliary functions $%
h_{s,t}^{1,x},h_{s,t}^{2},h_{s,t}^{3,x}:(0,\infty )\rightarrow (0,\infty )$
by 
\begin{align*}
\begin{split}
h_{s,t}^{1,x}(y)& =\lambda _{s,t}^{\ast }\left( \left( u_{s}^{\prime
}\right) ^{-1}\left( \lambda _{0,s}^{\ast }(x)\hat{N}_{0,s}^{\prime }\left(
1-w_{0,s}\left( F_{0,s}^{\rho }(y)\right) \right) \right) \right) , \\
h_{s,t}^{2}(y)& =\hat{N}_{s,t}^{\prime }\left( 1-w_{s,t}\left( F_{s,t}^{\rho
}(y)\right) \right) , \\
h_{s,t}^{3,x}(y)& =\lambda _{0,t}^{\ast }(x)\hat{N}_{0,t}^{\prime }\left(
1-w_{0,t}\left( F_{0,t}^{\rho }(y)\right) \right) .
\end{split}%
\end{align*}%
Since $\rho _{0,s}$ and $\rho _{s,t}$ are independent with $\rho _{0,s}\rho
_{s,t}=\rho _{0,t}$ we deduce that, for all $y,z>0,$  
\begin{align}
h_{s,t}^{1,x}(y)h_{s,t}^{2}(z)=h_{s,t}^{3,x}(yz)  .\label{eq:Relationh}
\end{align}%

Next, suppose that there exist $y_{1},y_{2}$ with $0<y_{1}<y_{2}$ and such
that $h_{s,t}^{1,x}(y_{1})=h_{s,t}^{1,x}(y_{2})$. Then, equality (\ref%
{eq:Relationh}) together with the monotonicity of $\widehat{N}^{\prime }$
imply that $h_{s,t}^{3,x}$, and in turn $h_{s,t}^{1,x}$ and $h_{s,t}^{2}$,
are constant. Hence, $\hat{N}_{s,t}^{\prime }(z)=1,$ for all $z\in \lbrack
0,1]$, which is the case if and only 
\begin{align*}
N_{s,t}(z)\leqslant z-1
\end{align*}%
for all $z\in \lbrack 0,1]$. 

Using the definition of $N_{s,t}$ in (\ref{eq:FunctionN}) and substituting $%
x=w_{s,t}^{-1}(1-z)$ yields that the above inequality is equivalent to (\ref%
{distortion-forward-degenerate}). The same argument can be made if there
exist $z_{1},z_{2}$ with $0<z_{1}<z_{2}$ and such that $%
h_{s,t}^{2}(z_{1})=h_{s,t}^{2}(z_{2})$. Similarly, if there exist $\xi
_{1},\xi _{2}$ with $0<\xi _{1}<\xi _{2}$ and $h_{s,t}^{3,x}(\xi
_{1})=h_{s,t}^{3,x}(\xi _{2})$, then $h_{s,t}^{1,x}(\xi
_{1})h_{s,t}^{2}(1)=h_{s,t}^{3,x}(\xi _{1})=h_{s,t}^{3,x}(\xi
_{2})=h_{s,t}^{1,x}(\xi _{2})h_{s,t}^{2}(1)$ and the statement follows.

Hence, there are two cases: either the family of probability distortion
functions satisfies (\ref{distortion-forward-degenerate}) or $h_{s,t}^{1,x}$
is strictly decreasing and $h_{s,t}^{2}$,$h_{s,t}^{3,x}$ strictly
increasing. On the other hand, the latter case in turn is equivalent to $%
N_{s,t}(\cdot )$ being concave, that is, that the probability distortion $%
w_{s,t}$ satisfies the Jin-Zhou monotonicity condition, namely, that
the function $\frac{\left( F_{s,t}^{\rho }\right) ^{-1}(\cdot )}{%
w_{s,t}^{\prime }(\cdot )}$ is nondecreasing.

Next, we suppose that the family of probability distortion functions does
satisfy the Jin-Zhou monotonicity condition. Because we have just shown that
in this case $N_{s,t}$ is concave, and thus coincides with its concave
envelope, a straightforward computation shows that the functions $%
h_{s,t}^{1,x},h_{s,t}^{2}$ and $h_{s,t}^{3,x}$ simplify to%
\begin{align*}
h_{s,t}^{1,x}(y)=\lambda _{s,t}^{\ast }\left( \left( u_{s}^{\prime }\right)
^{-1}\left( \lambda _{0,s}^{\ast }(x)\frac{y}{w_{0,s}^{\prime }\left(
F_{0,s}^{\rho }\left( y\right) \right) }\right) \right) ,
\end{align*}%
\begin{align*}
h_{s,t}^{2}(y)=\frac{y}{w_{s,t}^{\prime }\left( F_{s,t}^{\rho }\left(
y\right) \right) }\text{ \  \ and \  \  \ }h_{s,t}^{3,x}(y)=\lambda
_{0,t}^{\ast }(x)\frac{y}{w_{0,t}^{\prime }\left( F_{0,t}^{\rho }\left(
y\right) \right) }.
\end{align*}
Taking $y=1$ in (\ref{eq:Relationh}) yields $%
h_{s,t}^{3,x}(z)=h_{s,t}^{1,x}(1)h_{s,t}^{2}(z)$ while taking $z=1$ gives $%
h_{s,t}^{3,x}(y)=h_{s,t}^{1,x}(y)h_{s,t}^{2}(1)$. Combining the two gives 
\begin{align*}
h_{s,t}^{3,x}(yz)=h_{s,t}^{1,x}(y)h_{s,t}^{2}(z)=\frac{%
h_{s,t}^{3,x}(y)h_{s,t}^{3,x}(z)}{h_{s,t}^{1,x}(1)h_{s,t}^{2}(1)}=\frac{%
h_{s,t}^{3,x}(y)h_{s,t}^{3,x}(z)}{h_{s,t}^{3,x}(1)}.
\end{align*}

Next, we define $g:\mathbb{R}\rightarrow \mathbb{R}$ by $g(z):=\log \left(
h_{s,t}^{3,x}\left( e^{z}\right) \right) -\log h_{s,t}^{3,x}(1)$. We deduce
that $g$ satisfies Cauchy's functional equation $g(y+z)=g(y)+g(z)$.
Indeed, 
\begin{align*}
g(y+z)& =\log \left( h_{s,t}^{3,x}\left( e^{y+z}\right) \right) -\log
h_{s,t}^{3,x}(1) \\
& =\log \left( \frac{h_{s,t}^{3,x}\left( e^{y}\right) h_{s,t}^{3,x}\left(
e^{z}\right) }{h_{s,t}^{3,x}(1)}\right) -\log h_{s,t}^{3,x}(1) \\
& =\log \left( h_{s,t}^{3,x}\left( e^{y}\right) \right) +\log \left(
h_{s,t}^{3,x}\left( e^{z}\right) \right) -\log \left(
h_{s,t}^{3,x}(1)\right) -\log h_{s,t}^{3,x}(1) \\
& =g(y)+g(z),
\end{align*}%
for any $y,z\in \mathbb{R}$. Since $g$ is continuous there must be a $\gamma
\in \mathbb{R}$ such that $g(z)=\gamma z$, $z\in \mathbb{R}$. This, in turn,
yields that for $z>0,$ 
\begin{align*}
h_{s,t}^{3,x}(z)=h_{s,t}^{3,x}(1)z^{\gamma }.
\end{align*}

On the other hand, because $h_{s,t}^{3,x}$ is strictly increasing when the family of probability distortion functions does
satisfy the Jin-Zhou monotonicity condition, it must be that $\gamma $ is positive. We, therefore, obtain
that for $p\in \left[ 0,1\right] ,$ 
\begin{align*}
w_{s,t}^{\prime }(p)=\frac{\lambda _{s,t}^{\ast }(x)}{h_{s,t}^{3,x}(1)}%
\left( \left( F_{s,t}^{\rho }\right) ^{-1}(p)\right) ^{1-\gamma }.
\end{align*}%
Furthermore, since 
\begin{align*}
1=\int_{0}^{1}w_{s,t}^{\prime }(p)dp=\int_{0}^{1}\frac{\lambda _{s,t}^{\ast
}(x)}{h_{s,t}^{3,x}(1)}\left( \left( F_{s,t}^{\rho }\right) ^{-1}(p)\right)
^{1-\gamma }dp=\frac{\lambda _{s,t}^{\ast }(x)}{h_{s,t}^{3,x}(1)}\mathbb{E}%
\left[ \rho _{s,t}^{1-\gamma }\right] ,
\end{align*}%
we have that $h_{s,t}^{3,x}(1)=\lambda _{s,t}^{\ast }(x)\mathbb{E}\left[
\rho _{0,t}^{1-\gamma }\right] $ and in turn 
\begin{align*}
w_{s,t}(p)=\frac{1}{\mathbb{E}\left[ \rho _{s,t}^{1-\gamma }\right] }%
\int_{0}^{p}\left( \left( F_{s,t}^{\rho }\right) ^{-1}(q)\right) ^{1-\gamma
}dq.
\end{align*}%
Finally, since $h_{s,t}^{2}(z)=\frac{h_{s,t}^{3,x}(z)}{h_{s,t}^{1,x}(1)}$,
we obtain (\ref{distortion-forward}) for any $0\leqslant s<t$, using the same
arguments as above. \qed

\subsection{Proof of Corollary \protect \ref{cor:WeightingIndependentInitialTime}}

For the first case of Theorem \ref{thm:WeightingNecessary}, the only $\gamma
>0$ for which the probability distortion 
\begin{align*}
w_{s,t}(p)=\frac{1}{\mathbb{E}\left[ \rho _{s,t}^{1-\gamma }\right] }%
\int_{0}^{p}\left( \left( F_{s,t}^{\rho }\right) ^{-1}(q)\right) ^{1-\gamma
}dq
\end{align*}%
is independent of $s,$ for every $s\in \left[ 0,t\right) ,$ is when $\gamma =1$ and the assertion follows. 

The second case of Theorem \ref{thm:WeightingNecessary} cannot happen when $w_{s,t}$ is independent of $s$. Indeed, when $s$ goes to $t$, then the right-hand side of (\ref{distortion-forward-degenerate})  converges to the mapping $p\mapsto \mathbf{1}_{\left \{
p>0\right \} }$. However, the distortion function $w_{s,t}$ must be continuous, and we conclude.\qed

\subsection{Proof of Proposition \protect \ref{prop:OptimalWealthLagrangianMultiplier}}

Case ii) follows immediately from the proof of Theorem \ref{thm:WeightingNecessary} and thus we only need to show case i). We first show that the Lagrangian multiplier satisfies $\lambda _{s,t}^{\ast}(X_{s}^{\ast }) = u_{s}^{\prime}  \left( X_s^{\ast} \right) $. Indeed, from Theorem \ref%
{thm:WeightingNecessary} and the functional relation (\ref%
{eq:TCFunctionalEquation}), we deduce that  
\begin{align*}
\lambda _{s,t}^{\ast }\left( \left( u_{s}^{\prime }\right) ^{-1}\left(
\lambda _{0,s}^{\ast }(x)\mathbb{E}\left[ \rho _{0,s}^{1-\gamma }\right]
\rho _{0,s}^{\gamma }\right) \right) \mathbb{E}\left[ \rho _{s,t}^{1-\gamma }%
\right] \rho _{s,t}^{\gamma }=\lambda _{0,t}^{\ast }(x)\mathbb{E}\left[ \rho
_{0,t}^{1-\gamma }\right] \rho _{0,t}^{\gamma }.
\end{align*}%
Therefore, 
\begin{align*}
\lambda _{s,t}^{\ast }\left( \left( u_{s}^{\prime }\right) ^{-1}\left(
u_{0}^{\prime }(x)\mathbb{E}\left[ \rho _{0,s}^{1-\gamma }\right] \rho
_{0,s}^{\gamma }\right) \right) =u_{0}^{\prime }(x)\mathbb{E}\left[ \rho
_{0,s}^{1-\gamma }\right] \rho _{0,s}^{\gamma }.
\end{align*}%
From this, we conclude that $\lambda _{s,t}^{\ast }(x)=u_{s}^{\prime }(x)$,
for all $x>0$.

Next, we prove that the optimal wealth process is given by (\ref
{optimal-wealth-general}). Because the distortion functions in (\ref{distortion-forward}) satisfy the Jin-Zhou monotonicity condition, the optimal wealth process is given by 
\begin{align*}
X_{t}^{\ast }= \left(u_{t}^{\prime} \right)^{-1}\left( \lambda _{0,t}^{\ast }(x)\frac{\rho
_{t}}{w_{0,t}^{\prime \rho }{}_{t}(\rho _{t}))}\right) =\left( u_{t}^{\prime
}\right) ^{-1}\left( u_{0}^{\prime }(x)\mathbb{E}\left[ \rho _{t}^{1-\gamma }%
\right] \rho _{t}^{\gamma }\right) ,\quad t\geqslant 0,
\end{align*}%
where we used the form of the Lagrangian multiplier determined above. From this, we deduce that 
\begin{align*}
X_{t}^{\ast }=\left( u_{t}^{\prime }\right) ^{-1}\left( \lambda _{s,t}^{\ast
}\left( X_{s}^{\ast }\right) \mathbb{E}\left[ \rho _{s,t}^{1-\gamma }\right]
\rho _{s,t}^{\gamma }\right) =\left( u_{t}^{\prime }\right) ^{-1}\left(
u_{0}^{\prime }(x)\mathbb{E}\left[ \rho _{t}^{1-\gamma }\right] \rho
_{t}^{\gamma }\right) ,\quad 0\leqslant s<t.
\end{align*}%
\qed

\subsection{Proof of Proposition \protect \ref{prop:Pessimism}}

Note that $w (p) \geq p$ for all $p \in [0,1]$ if and only if $\gamma \geqslant 1$ and $w_1 (p) \geq w_2 (p)$ for all $p \in [0,1]$ if and only if $\gamma_1 \geq \gamma_2$. Thus, the assertion follows by Propositions 2.3 and 2.6 in \cite{GhossoubHeXX}. \qed

\subsection{Proof of Theorem \protect \ref{thm:ClassicalForwardUtilityEquivalentMeasure}}

We first prove the correspondence between forward rank-dependent performance processes and deterministic, time-monotone forward performance processes under the distorted probability measure. We start with the converse direction.

Let $\left( \left( u_{t}\right) _{t\geqslant 0},\left( w_{s,t}\right) _{0\leqslant
s<t}\right) $ be a forward rank-dependent performance criterion. If (\ref{distortion-forward-degenerate}) holds and, consequently, the optimal wealth process 
$X^{\ast }=\left( X_{t}^{\ast }\right) _{t\geqslant 0}$ is constant, we set $\gamma =0$. Then, the risky assets in the $\gamma $-distorted market become martingales and do not offer any excess return under the
distorted measure. It is thus optimal to invest everything into the
risk-free asset. Thus, $u_t (x) = u_0(x) = v(x,0)$ is a deterministic, time-monotone forward performance criterion in the $\gamma$-distorted market by Proposition \ref{prop:DefinitionForwardEquivalent} and Proposition \ref{prop:TimeConsistency}. 

Now suppose that we are in case (\ref{distortion-forward}). 
Let $\mathbb{P}_{\gamma }$ be given by (\ref{gamma-measure}) and denote the expectation under it by $\mathbb{E}_{\gamma }$. According
to Proposition \ref{prop:OptimalWealthLagrangianMultiplier}, the wealth
process $X^{\ast }$
also solves the family of expected utility maximization problems generated by the optimal policy, say $\pi^{*}$,

\begin{align}\label{prob:EUDistortedMeasure}
\max_{X}\mathbb{E}_{\gamma }\left[ u_{t}\left( X\right) \right]   
\end{align}
with $\mathbb{E}_{\gamma }\left[ \left. \tilde{\rho}_{s,t}X\right \vert \mathcal{F}_{s}\right] =X_{s}^{\ast }, \enskip X\geqslant 0,
\enskip X\in \mathcal{F}_{t}$.
Moreover, we have that 
\begin{align}
\begin{split}\label{eq:RDUValueEqualsValueDistortedMeasure}
u_{s}(X_{s}^{\ast })& =\int_{0}^{\infty }u_{t}(\xi )d\left( -w_{s,t}\left(
1-F_{X_{t}^{\ast }|\mathcal{F}_{s}}(\xi )\right) \right)  \\
& =\int_{0}^{\infty }u_{t}(\xi )d\left( -w_{s,t}\left( \mathbb{P}\left[
\left. \rho _{s,t}\leqslant \left( \frac{u_{t}^{\prime }(\xi )}{u_{s}^{\prime
}\left( X_{s}^{\ast }\right) \mathbb{E}\left[ \rho _{s,t}^{1-\gamma }\right] 
}\right) ^{1/\gamma }\right \vert \mathcal{F}_{s}\right] \right) \right)  \\
& =\int_{0}^{\infty }u_{t}\left( \left( u_{t}^{\prime }\right) ^{-1}\left(
u_{s}^{\prime }\left( X_{s}^{\ast }\right) \mathbb{E}\left[ \rho
_{s,t}^{1-\gamma }\right] y^{\gamma }\right) \right) dw_{s,t}\left(
F_{s,t}^{\rho }(y)\right)  \\
& =\mathbb{E}\left[ \left. u_{t}\left( \left( u_{t}^{\prime }\right)
^{-1}\left( u_{s}^{\prime }\left( X_{s}^{\ast }\right) \mathbb{E}\left[ \rho
_{s,t}^{1-\gamma }\right] \rho _{s,t}^{\gamma }\right) \right)
w_{s,t}^{\prime }\left( F_{s,t}^{\rho }\left( \rho _{s,t}\right) \right)
\right \vert \mathcal{F}_{s}\right]  \\
& =\mathbb{E}\left[ \left. u_{t}\left( \left( u_{t}^{\prime }\right)
^{-1}\left( u_{s}^{\prime }\left( X_{s}^{\ast }\right) \widetilde{\rho }%
_{s,t}\right) \right) \frac{\rho _{s,t}^{1-\gamma }}{\mathbb{E}\left[ \rho
_{s,t}^{1-\gamma }\right] }\right \vert \mathcal{F}_{s}\right]  \\
& =\mathbb{E}_{\gamma }\left[ \left. u_{t}\left( \left( u_{t}^{\prime
}\right) ^{-1}\left( u_{s}^{\prime }\left( X_{s}^{\ast }\right) \widetilde{%
\rho }_{s,t}\right) \right) \right \vert \mathcal{F}_{s}\right]  \\
& =\mathbb{E}_{\gamma }\left[ \left. u_{t}\left( X_{t}^{\ast }\right)
\right \vert \mathcal{F}_{s}\right] .
\end{split}
\end{align}
Therefore, for each fixed $t\geqslant 0$, we have that, from the one hand, $u_{s}$
corresponds to the value function of the expected utility maximization
problem under the distorted measure $\mathbb{P}_{\gamma }$, (\ref%
{prob:EUDistortedMeasure}), with time horizon $t$ and utility function $u_{t}$, and, from the other, this policy $\pi^{*}$ is optimal. Hence, $\mathbb{E}_{\gamma }\left[ \left. u_{t}\left( X_{t}^{\pi }\right)
\right \vert \mathcal{F}_{s}\right] \leqslant u_{s}\left( X_{s}^{\pi }\right) $
for any admissible policy $\pi $ with the same argument as in Proposition %
\ref{prop:TimeConsistency}. Thus, $\left( u_t \right)_{t \geqslant 0}$ is a forward performance criterion in the $\gamma$-distorted market.

To establish the other direction we work as follows. Let $\gamma \geqslant 0$ and let $\left( u_{t}\right) _{t\geqslant 0}$ be a deterministic, time-monotone forward performance process in the $\gamma$-distorted market. Together with $\tilde{w}_{s,t}(p)\equiv p$, $p\in \lbrack 0,1]$ for all $0\leqslant s<t$, the pair $\left( \left( u_{t}\right)_{t\geqslant 0},\left( \tilde{w}_{s,t}\right)_{0\leqslant s<t}\right)$ is a forward rank-dependent performance process in the $\gamma$-distorted market according to Proposition \ref{prop:DefinitionForwardEquivalent}. According to Proposition \ref{prop:OptimalWealthLagrangianMultiplier}, the corresponding optimal wealth process is given by
\begin{align*}
X_{t}^{\ast } = \left( u_{t}^{\prime }\right) ^{-1}\left(
u_{0}^{\prime }(x) \rho
_{\gamma, t} \right)  =\left( u_{t}^{\prime }\right) ^{-1}\left(
u_{0}^{\prime }(x)\mathbb{E}\left[ \rho _{t}^{1-\gamma }\right] \rho
_{t}^{\gamma }\right) ,\quad 0\leqslant t.  
\end{align*}
Next, we define the family of probability distortions $\left(
w_{s,t}\right) _{0\leqslant s<t}$ by (\ref{distortion-forward}) and note that the optimal wealth process $X^{\ast}$ solves (\ref{prob:RDUIntermediateTime}) for any $0 \leqslant s < t$. Therefore, the pair $\left( \left( u_{t}\right) _{t\geqslant 0},\left(
w_{s,t}\right) _{0\leqslant s<t}\right) $ is a time-consistent
rank-dependent performance criterion and because of (\ref{eq:RDUValueEqualsValueDistortedMeasure}) also preserves the performance value. Hence, it is a forward rank-dependent performance criterion according to Proposition \ref{prop:TimeConsistency}.

Finally, we deduce the construction method for forward rank-dependent performance criteria. Using Girsanov's
theorem, it is straightforward to compute that the market price of risk
under the distorted probability measure $\mathbb{P}_{\gamma }$ is given by $%
\gamma \lambda $. The statement then follows as a direct consequence of
the results on time-monotone forward criteria in \cite{MusielaZariphopoulou10a} and the correspondence
shown above.
\qed

\subsection{Proof of Proposition \protect \ref{prop:ConstructedOptimalWealthStrategy}}
Following the results in \cite{MusielaZariphopoulou10a}, we deduce that the optimal wealth under the time-monotone
forward performance criteria in the $\gamma $-distorted market is given by
\begin{align*}
X_{t}^{\ast }=h\left( h^{-1}\left( 0,t\right) +\int_{0}^{t}\Vert \lambda
_{\gamma ,s}\Vert ^{2}ds+\int_{0}^{t}\lambda _{\gamma ,s}dW_{\gamma
,s},\int_{0}^{t}\Vert \lambda _{\gamma ,s}\Vert ^{2}ds\right) ,
\end{align*}
with $h$ as in (\ref{h-function}). Using (\ref{gamma-MPR})
and (\ref{W-gamma}) we conclude. We similarly deduce (\ref{pi-gamma}).
\qed

\subsection{Proof of Theorem \protect \ref{thm:DynamicRDU}}

Recall that the optimal solution to 
\begin{align*}
\begin{split}
\max_{X}& \quad \int_{0}^{\infty }u_{t,T}(\xi )d\left( -w_{t,T}\left(
1-F_{X}(\xi )\right) \right)  \\
\mathrm{s.t.}& \quad \mathbb{E}\left[ \rho _{t,T}X|\mathcal{F}_{t}\right] =%
\mathbb{E}\left[ \rho _{t,T}X^{\ast }|\mathcal{F}_{t}\right] ,\enskip X\geqslant
0,\enskip X\enskip \mathrm{is}\enskip \mathcal{F}_{T}-\mathrm{measurable}.
\end{split}%
\end{align*}%
is given by 
\begin{align*}
X^{\ast ,t}=(u_{t,T}^{\prime })^{-1}\left( \lambda _{t,T}^{\ast }\left( 
\mathbb{E}\left[ \rho _{t,T}X^{\ast }|\mathcal{F}_{t}\right] \right) \hat{N}%
_{t,T}^{\prime }\left( 1-w_{t,T}\left( F_{t,T}^{\rho }(\rho _{t,T})\right)
\right) \right) ,
\end{align*}%
where $\hat{N}_{t,T}$ is the concave envelope of (\ref{eq:FunctionN}) and $%
\lambda _{t,T}^{\ast }(X_{t})>0$ is such that $\mathbb{E}\left[ \rho
_{t,T}X^{\ast ,t}\big \vert \mathcal{F}_{t}\right] =\mathbb{E}\left[ \rho
_{T}X^{\ast }|\mathcal{F}_{t}\right] $. Optimality of the initial optimal
solution $X^{0,\ast }=X^{\ast }$ is thus maintained if and only if 
\begin{align}\label{eq:OptimalityInitialSolutionMaintained}
\begin{split}
\lambda _{t,T}^{\ast }& \left( \mathbb{E}\left[ \rho _{t,T}X^{\ast }|%
\mathcal{F}_{t}\right] \right) \hat{N}_{t,T}^{\prime }\left( 1-w_{t,T}\left(
F_{t,T}^{\rho }(\rho _{t,T})\right) \right)  \\
& =u_{t,T}^{\prime }\left( (u_{0,T}^{\prime })^{-1}\left( \lambda
_{0,T}^{\ast }\left( x\right) \hat{N}_{0,T}^{\prime }\left( 1-w_{0,T}\left(
F_{0,T}^{\rho }(\rho _{0,T})\right) \right) \right) \right) .
\end{split}%
\end{align}
Similar to the proof of Theorem \ref{thm:WeightingNecessary}, we define $
g_{t}^{1,x},g_{t}^{2},g_{t}^{3,x}:(0,\infty )\rightarrow (0,\infty )$ by 
\begin{align}\label{eq:DefinitionG}
\begin{split}
g_{t}^{1,x}(y)& =\lambda _{t,T}^{\ast }\left( \mathbb{E}\left[ \rho
_{t,T}(u_{0,T}^{\prime })^{-1}\left( \lambda _{0,T}^{\ast }\left( x\right) 
\hat{N}_{0,T}^{\prime }\left( 1-w_{0,T}\left( F_{0,T}^{\rho }(y\rho
_{t,T})\right) \right) \right) \right] \right) , \\
g_{t}^{2}(y)& =\hat{N}_{t,T}^{\prime }\left( 1-w_{t,T}\left( F_{t,T}^{\rho
}(y)\right) \right) , \\
g_{t}^{3,x}(y)& =u_{t,T}^{\prime }\left( (u_{0,T}^{\prime })^{-1}\left(
\lambda _{0,T}^{\ast }\left( x\right) \hat{N}_{0,T}^{\prime }\left(
1-w_{0,T}\left( F_{0,T}^{\rho }(y)\right) \right) \right) \right) .
\end{split}%
\end{align}
By the independence of $\rho _{0,t}$ and $\rho _{t,T}$ and since $\rho
_{0,t}\rho _{t,T}=\rho _{0,T}$ we have (\ref{eq:OptimalityInitialSolutionMaintained}) if and only if 
\begin{align}\label{eq:RelationG}
g_{t}^{1,x}(y)g_{t}^{2}(z)=g_{t}^{3,x}(yz)
\end{align}%
for all $y,z>0$. Since $u_{t,T}^{\prime }$ is strictly decreasing for any $t\in \lbrack 0,T)$, we can make the same argument as in the proof of Theorem %
\ref{thm:WeightingNecessary} to conclude that this holds if and only if
either 
\begin{align*}
w_{t,T}(p)\geqslant \mathbb{E}\left[ \rho _{t,T}\boldsymbol{1}_{\left \{ \rho
_{t,T}\leqslant \left( F_{t,T}^{\rho }\right) ^{-1}(p)\right \} }\right] ,\quad
p\in \lbrack 0,1],
\end{align*}%
for any $t\in \lbrack 0,T)$, or the family of probability distortions
functions satisfies the Jin-Zhou monotonicity condition. The first case
proves the first part of the theorem.

For the latter case, we first show the only if direction by following a
similar line of argument as in Theorem \ref{thm:WeightingNecessary}. If the
probability distortion functions satisfy the Jin-Zhou monotonicity
condition, $g^{1,x}_{t}, g^{2}_{t}, g^{3,x}_{t}: (0, \infty) \rightarrow (0,
\infty)$ defined in (\ref{eq:DefinitionG}) simplify to 
\begin{align*}
g^{1,x}_{t} (y) & = \lambda^*_{t,T} \left( \mathbb{E}\left[ \rho_{t,T}
(u^{\prime }_{0,T})^{-1} \left( \lambda^*_{0,T} \left( x \right) \frac{y
\rho_{t,T}}{w_{0,T}^{\prime }\left( F^{\rho}_{0,T} \left( y \rho_{t,T}
\right) \right)} \right) \right] \right), \\
g^{2}_{t} (y) & = \frac{y}{w_{t,T}^{\prime }\left(F^\rho_{t,T} ( y) \right)},
\\
g^{3,x}_{t} (y) & = u^{\prime }_{t,T} \left( (u^{\prime }_{0,T})^{-1} \left(
\lambda^*_{0,T} \left( x \right) \frac{y}{w_{0,T}^{\prime
}\left(F^\rho_{0,T} (y) \right)} \right) \right).
\end{align*}
As in the proof of Theorem \ref{thm:WeightingNecessary}, we can show that $%
g^{3,x}_{t}$ satisfies Cauchy's functional equation and conclude that there
is a $\gamma_t > 0$ such that $g^{3,x}_{t} (y) = g^{3,x}_{t} (1)
y^{\gamma_t} $ for all $y > 0$. Thus, by virtue of (\ref{eq:RelationG}), 
\begin{align*}
\frac{g^{1,x}_{t}(y)}{y^{\gamma_t}} = g^{3,x}_{t} (1) \frac{z^{\gamma_t}}{%
g^{2}_{t} (z)} = C_{x,t}.
\end{align*}
for some constant $C_{x,t}$. From the definition of $g^{2}_{t}$ we obtain
that for any $t \in (0, T)$, 
\begin{align*}
w_{t,T} (p) = \frac{1}{\mathbb{E}\left[ \rho_{t,T}^{1-\gamma_t}\right]}
\int_0^p \left( \left( F^\rho_{t,T} \right)^{-1} (q) \right)^{1-\gamma_t} dq
.
\end{align*}
By continuity of $w_{t,T}$ in $t$ at zero, 
\begin{align*}
w_{0,T} (p) = \frac{1}{\mathbb{E}\left[ \rho_{0,T}^{1-\gamma_0}%
\right]} \int_0^p \left( \left( F^\rho_{0,T} \right)^{-1} (q)
\right)^{1-\gamma_0} dq
\end{align*}
where $\gamma_0 = \lim_{t\searrow 0} \gamma_t$. From $g^{3,x}_{t} (y) =
g^{3,x}_{t} (1) y^{\gamma_t}$ we thus obtain that 
\begin{align*}
u^{\prime }_{t,T} \left( (u^{\prime }_{0,T})^{-1} \left( \lambda^*_{0,T}
\left( x \right) \mathbb{E}\left[ \rho_{0,T}^{1-\gamma_0} \right]
y^{\gamma_0} \right) \right) = g^{3,x}_{t} (1) y^{\gamma_t}.
\end{align*}
With the substitution $z = \left( u_{0,T}^{\prime }\right)^{-1} \left(
\lambda^*_{0,T} (x) \mathbb{E}\left[ \rho_{0,T}^{1-\gamma_0} \right]
y^{\gamma_0} \right)$ this becomes 
\begin{align}  \label{eq:RelationDerivativeUandW}
u_{t,T}^{\prime }\left( z \right) = \frac{g^{3,x}_{t} (1) }{ \left(
\lambda^*_{0,T} (x) \mathbb{E}\left[ \rho_{0,T}^{1-\gamma_0} \right]
\right)^{\gamma_t / \gamma_0}} \left( u_{0,T}^{\prime }(z) \right)^{\gamma_t
/ \gamma_0} .
\end{align}
Differentiating (\ref{eq:RelationDerivativeUandW}) with respect to $z$
yields 
\begin{align}  \label{eq:RelationSecondDerivativeUandW}
u_{t,T}^{\prime \prime }\left( z \right) = \frac{g^{3,x}_{t} (1) }{ \left(
\lambda^*_{0,T} (x) \mathbb{E}\left[ \rho_{0,T}^{1-\gamma_0} \right]
\right)^{\gamma_t / \gamma_0}} \frac{\gamma_t}{\gamma_0} \left(
u_{0,T}^{\prime }(z) \right)^{\gamma_t / \gamma_0 -1} u_{0,T}^{\prime \prime
}(z).
\end{align}
Dividing (\ref{eq:RelationSecondDerivativeUandW}) by (\ref%
{eq:RelationDerivativeUandW}) gives (\ref{eq:RiskAversionDynamicRDU}). 
\newline

For the if direction, we fix $t\in (0,T)$ and first note that (\ref%
{eq:RiskAversionDynamicRDU}) is equivalent to 
\begin{align*}
\frac{d}{dz}\log \left( u_{t,T}^{\prime }(z)\right) =\frac{\gamma _{t}}{%
\gamma _{0}}\frac{d}{dz}\log \left( u_{0,T}^{\prime }(z)\right) 
\end{align*}%
and thus $u_{t,T}^{\prime }(z)=\widetilde{C}\left( u_{0,T}^{\prime
}(z)\right) ^{\gamma _{t}/\gamma _{0}}$ for some constant $\widetilde{C}\in 
\mathbb{R}$. From this we derive 
\begin{align}\label{eq:DynamicInverseMarginalRelation}
\left( u_{t,T}^{\prime }\right) ^{-1}\left( \widetilde{C}z^{\gamma
_{t}}\right) =\left( u_{0,T}^{\prime }\right) ^{-1}\left( z^{\gamma
_{0}}\right) ,
\end{align}
$z\in (0,\infty )$. According to Theorem \ref{thm:BackwardRDEUSolution}, the
optimal solution to (\ref{prob:DynamicRDU}) is given by 
\begin{align*}
X^{\ast ,t}=(u_{t,T}^{\prime })^{-1}\left( \lambda _{t,T}^{\ast }\left( 
\mathbb{E}\left[ \rho _{t,T}X^{\ast }|\mathcal{F}_{t}\right] \right) \mathbb{%
E}\left[ \rho _{t,T}^{1-\gamma _{t}}\right] \rho _{t,T}^{\gamma _{t}}\right)
,
\end{align*}%
where $\lambda _{t,T}^{\ast }(X_{t})>0$ is such that $\mathbb{E}\left[
\left. \rho _{t,T}X^{\ast ,t}\right \vert \mathcal{F}_{t}\right] =\mathbb{E}%
\left[ \rho _{T}X^{\ast }|\mathcal{F}_{t}\right] $. From Theorem 4.1 in \cite{HeKouwenbergZhou17} we furthermore have that $\mathbb{E}\left[ \rho
_{T}X^{\ast }|\mathcal{F}_{t}\right] =G_{t,\lambda ^{\ast }(x)}\left( \rho
_{t}\right) $ for some strictly decreasing and thus invertible function $%
G_{t,\lambda ^{\ast }(x)}$. We define $\lambda _{t,T}:(0,\infty )\rightarrow
(0,\infty )$ by 
\begin{align*}
\lambda _{t,T}(\xi ):=\widetilde{C}\left( \lambda _{0,T}^{\ast }(x)\right)
^{\gamma _{t}/\gamma _{0}}\frac{\mathbb{E}\left[ \rho _{0,T}^{1-\gamma _{0}}%
\right] ^{\gamma _{t}/\gamma _{0}}}{\mathbb{E}\left[ \rho _{t,T}^{1-\gamma
_{t}}\right] }G_{t,\lambda ^{\ast }(x)}^{-1}(\xi )^{\gamma _{t}}.
\end{align*}%
Using the relation (\ref{eq:DynamicInverseMarginalRelation}) we obtain 
\begin{align*}
(u_{t,T}^{\prime })^{-1}\left( \lambda _{t,T}\left( \mathbb{E}\left[ \rho
_{t,T}X^{\ast }|\mathcal{F}_{t}\right] \right) \mathbb{E}\left[ \rho
_{t,T}^{1-\gamma _{t}}\right] \rho _{t,T}^{\gamma _{t}}\right) &
=(u_{t,T}^{\prime })^{-1}\left( \widetilde{C}\left( \lambda _{0,T}^{\ast
}(x)\right) ^{\frac{\gamma _{t}}{\gamma _{0}}}\mathbb{E}\left[ \rho
_{0,T}^{1-\gamma _{0}}\right] ^{\frac{\gamma _{t}}{\gamma _{0}}}\rho
_{0,T}^{\gamma _{t}}\right)  \\
& =(u_{0,T}^{\prime })^{-1}\left( \lambda _{0,T}^{\ast }(x)\mathbb{E}\left[
\rho _{0,T}^{1-\gamma _{0}}\right] \rho _{0,T}^{\gamma _{0}}\right)  \\
& =X^{\ast }.
\end{align*}%
We in particular have that 
\begin{align*}
\mathbb{E}\left[ \rho _{t,T}(u_{t,T}^{\prime })^{-1}\left( \lambda
_{t,T}\left( \mathbb{E}\left[ \rho _{t,T}X^{\ast }|\mathcal{F}_{t}\right]
\right) \mathbb{E}\left[ \rho _{t,T}^{1-\gamma _{t}}\right] \rho
_{t,T}^{\gamma _{t}}\right) \bigg \vert \mathcal{F}_{t}\right] =\mathbb{E}%
\left[ \rho _{t,T}X^{\ast }\big \vert \mathcal{F}_{t}\right] .
\end{align*}%
By the uniqueness of the Lagrangian multiplier we conclude that $\lambda
_{t,T}(\cdot )=\lambda _{t,T}^{\ast }(\cdot )$ and therefore $X^{\ast
,t}=X^{\ast }$. \qed

\end{document}